\documentclass[fleqn,usenatbib]{mnras}

\usepackage{natbib}
\usepackage{newtxtext,newtxmath}
\usepackage{hyperref} 
\usepackage{subcaption}

\usepackage[T1]{fontenc}

\DeclareRobustCommand{\VAN}[3]{#2}
\let\VANthebibliography\thebibliography
\def\thebibliography{\DeclareRobustCommand{\VAN}[3]{##3}\VANthebibliography}

\usepackage{graphicx}	
\usepackage{amsmath}	
\usepackage{siunitx}
\usepackage{caption}
\usepackage{nicematrix}
\usepackage{color}

\graphicspath{{/img}}

\DeclareSIUnit\angstrom{\text{Å}}

\title[JWST/NIRSpec spectroscopy of intermediate-mass quiescent galaxies at $z \sim 3\texttt{--}4$]{JWST/NIRSpec spectroscopy of intermediate-mass quiescent galaxies at $z \sim 3\texttt{--}4$}

\author[R.A. Sato et al.]{
Riku A. Sato,$^{1}$\thanks{E-mail: sato.riku2019@toki.waseda.jp}
Akio K. Inoue,$^{1,2}$
Yuichi Harikane,$^{3}$
Rhythm Shimakawa,$^{4,5}$
Yuma Sugahara,$^{1,2,6}$
\newauthor
Yoichi Tamura,$^{7}$
Takuya Hashimoto,$^{8,9}$
Kei Ito,$^{10}$
Satoshi Yamanaka,$^{11}$
Ken Mawatari,$^{2}$
\newauthor
Yoshinobu Fudamoto,$^{12}$
Yi W. Ren$^{1}$
\\
$^{1}$Department of Pure and Applied Physics, Graduate School of Advanced Science and Engineering, \\ Faculty of Science and Engineering, Waseda University, 3-4-1 Okubo, Shinjuku, Tokyo 169-8555, Japan\\
$^{2}$Waseda research Institute of Science and Engineering, Faculty of Science and Engineering, Waseda University,\\ 3-4-1 Okubo, Shinjuku, Tokyo 169-8555, Japan\\
$^{3}$Institute for Cosmic Ray Research, The University of Tokyo, 5-1-5 Kashiwanoha, Kashiwa, Chiba 277-8582, Japan\\
$^{4}$Waseda Institute for Advanced Study (WIAS), Waseda University, 1-21-1, Nishi-Waseda, Shinjuku, Tokyo 169-0051, Japan\\
$^{5}$Center for Data Science, Waseda University, 1-6-1, Nishi-Waseda, Shinjuku, Tokyo 169-0051, Japan\\
$^{6}$National Astronomical Observatory of Japan, 2-21-1 Osawa, Mitaka, Tokyo 181–8588, Japan\\
$^{7}$Department of Physics, Graduate School of Science, Nagoya University, Furo, Chikusa, Nagoya, Aichi 464-8602, Japan\\
$^{8}$Division of Physics, Faculty of Pure and Applied Sciences, University of Tsukuba, Tsukuba, Ibaraki 305-8571, Japan\\
$^{9}$Tomonaga Center for the History of the Universe (TCHoU), Faculty of Pure and Applied Sciences, University of Tsukuba, Tsukuba, Ibaraki 305-8571, Japan\\
$^{10}$Department of Astronomy, Graduate School of Science, The University of Tokyo, 7-3-1 Hongo, Bunkyo-ku, Tokyo 113-0033, Japan\\
$^{11}$General Education Department, National Institute of Technology, Toba College, 1-1, Ikegami-cho, Toba, Mie 517-8501, Japan\\
$^{12}$Center for Frontier Science, Chiba University, 1-33 Yayoi-cho, Inage-ku, Chiba 263-8522, Japan
}

\date{Accepted XXX. Received YYY; in original form ZZZ}

\pubyear{2024}

\begin{document}
\label{firstpage}
\pagerange{\pageref{firstpage}--\pageref{lastpage}}
\maketitle

\begin{abstract}

We present the analysis of three intermediate-mass quiescent galaxies (QGs) with stellar masses of $\sim10^{10}M_{\rm \odot}$ at redshifts $z\sim 3\texttt{--}4$ using NIRSpec low-resolution spectroscopy. Utilising the SED fitting code \textit{BAGPIPES}, we confirm these target galaxies are consistent with quiescent population, with their specific star formation rates (sSFR) falling below 2-dex the star-forming main sequence at the same redshifts. Additionally, we identify these QGs to be less massive than those discovered in previous works, particularly prior to the JWST era. Two of our target galaxies exhibit the potentially-blended H${\alpha}$+[NII] emission line within their spectra with $S/N>5$. We discuss whether this feature comes from an Active Galactic Nucleus (AGN) or star formation although future high-resolution spectroscopy is required to reach a conclusion. One of the target galaxies is covered by JWST/NIRCam imaging of the \textit{PRIMER} survey. Using the 2D profile fitting code \textit{Galfit}, we examine its morphology, revealing a disc-like profile with a S\'{e}rsic index of $n=1.1 \pm 0.1$. On the size--mass relation, we find a potential distinction between less-massive ($\log_{10}{(M_*/M_\odot)}<10.3$) and massive ($\log_{10}{(M_*/M_\odot)}>10.3$)  QGs in their evolutionary pathways. The derived quenching timescales for our targets are less than $1~\si{Gyr}$. This may result from these galaxies being quenched by AGN feedback, supporting the AGN scenario of the emission line features.

\end{abstract}




\section{Introduction}

Quiescent galaxies (QGs) are passively evolving galaxies that do not display significant star formation activity when compared to the star formation main sequence \citep[SFMS; see e.g.][]{Brinchmann04,Whitaker12,Whitaker14,Speagle14,Santini17,Houston23,Popesso23}. 
Previous research has sought to understand the processes through which these galaxies cease their star formation activities. The cessation of star formation is commonly referred to as "quenching" in the context of galaxy evolution. The high-redshift quiescent population is regarded as the progenitors of local massive early-type galaxies (ETGs). Studying high-redshift QGs is crucial for understanding the formation of galaxies across cosmic time.

For years, several quenching mechanisms, such as feedback of Active Galactic Nuclei
(AGNs) or star formation \citep[e.g.][]{Croton06,Maiolino12,Peng15,Zolotov15,Piotrowska22,Bluck23b}, environmental quenching including ram pressure stripping \citep[e.g.][]{Boselli22}, and morphological quenching \citep[e.g.][]{Martig09} have been considered to explain the quenching mechanism \citep[e.g.][]{Gabor10,YJPeng10,Somerville15,Zolotov15,Bluck23b}.

The size-mass relation, which is between effective radius and stellar mass of galaxies, is important for understanding how galaxies are formed and evolved across cosmic time. QGs are more compact than star-forming galaxies (SFGs), and have larger size with larger stellar mass \citep[e.g.][]{vanderWel14,Nedkova21}. The size-mass relation for QGs has a steeper slope compared to that for SFGs \citep[e.g.][]{vanderWel14, Nedkova21}. 
In the context of the size evolution of QGs \citep[e.g.,][]{vanderWel14, Nedkova21}, minor mergers are believed to play a dominant role, as supported by past observations \citep[e.g.,][]{Bezanson09, Newman12, vanderWel14} and simulations \citep[e.g.,][]{Naab09}. These minor mergers contribute to the steeper size-mass relation observed for QGs \citep[e.g.,][]{Ownsworth14, vanDokkum15}. Some studies show the flatten of size-mass relation at the low-mass end \citep[e.g.][]{Kawinwanichakij21,Nedkova21}. Size evolution of galaxies might not be driven by mergers but rather by star formation at low-mass end \citep[e.g.][]{vanDokkum15, Cutler23}

QGs in the high-z universe are intrinsically so faint that it is challenging to detect them. Previously, NIR instruments on ground based telescopes made advancement of spectroscopic confirmation of QGs at $z>3$ \citep[][]{Glazebrook17,Schreiber18a,Schreiber18b,Tanaka19, Forrest20, Valentino20, D'Eugenio21, Antwi-Danso23, Kakimoto24}. However, these QGs found to be massive galaxies with their stellar mass of $\log_{10}{(M_*/M_\odot)}>10.3$. Photometric candidates at $z>3$ with stellar masses $\gtrsim10^9 M_\odot$ were also reported before JWST era \citep[e.g.][]{Davidzon17, Mawatari16, Merlin19, Santini19, Mawatari20}, but it is important to emphasise that these were not spectroscopically confirmed. The James Webb Space Telescope (JWST), with its remarkable sensitivity in the near infrared, has enabled the detection of these high-redshift, less-massive QGs, opening a new era of QG research. In recent studies, less-massive and/or high redshift ($z \gtrsim 3$) QGs have been detected using JWST, photometrically \citep[][]{Carnall23a, Valentino23, Alberts23, Cutler23}, and spectroscopically  \citep[][]{Carnall23b,Looser24,Marchesini23,Sandles23,Strait23,Nanayakkara24, Carnall24, deGraaff24,Weibel24ax} . 

Here we report three QGs at $z\simeq 3\texttt{--} 4$ observed by using JWST Near-Infrared Spectrograph (JWST/NIRSpec). We analyse the obtained low resolution ($R\sim100$) spectroscopic data. Our samples are less-massive galaxies with the stellar masses of $\log_{10}{(M_*/M_\odot)}<10.3$ estimated from spectral fitting. Two of our samples, HD1 and HD3, shows line emission feature considered as the H${\alpha}$+[NII] emission line. Additionally, HD3 has JWST Near-Infrared Camera (JWST/NIRCam) and JWST Mid-Infrared Instrument (JWST/MIRI) imaging data, allowing us to study the morphology of the low mass QG. 

This paper is structured as follows. In Section~\ref{sec:data}, we detail the data employed in this study. In Section~\ref{sec:analysis}, we explain our analysis methods and the results. In Section~\ref{sec:discussion}, we discuss about formation and evolution of QGs based on our results. Section~\ref{sec:conclusion} offers a summary of our findings.
In this paper, we adopt a standard set of the cosmological parameters of $(\Omega_m,\Omega_{\Lambda},h)=(0.3,0.7,0.7)$.

\section{Data}\label{sec:data}

\subsection{Target galaxies}

Target galaxies, named HD1, HD2, and HD3, have been identified as $H$-dropout galaxies in the wide and deep imaging data set in the COSMOS \citep{Scoville07} and SXDF \citep{Furusawa08} fields collected by using ground-based telescopes and Spitzer \citep[][]{Harikane22a}. \citet{Harikane22a} employed $grizy$ images from the Hyper Suprime-Cam Subaru Strategic Program (HSC-SSP) survey \citep{Aihara18, Aihara19} public data release 2 (PDR2), $JHK_s/K$ images from UltraVISTA DR4 \citep{McCracken12} and UKIDSS UDS DR11 \citep{Lawrence07}, and Spitzer/IRAC [3.6] and [4.5] images from the Spitzer Large Area Survey with Hyper-Suprime-Cam (SPLASH). While all of our targets are not Lyman Break Galaxies (LBGs) at $z>13$ initially expected, they have collectively revealed an intriguing population of QGs in the intermediate redshift range $z \sim 3 \text{--} 4$ as found by \citet{Harikane24},  which aligns with an alternative possibility suggested in \citet{Harikane22a}. Despite this deviation from their initial expectation, we pursued a comprehensive study of these galaxies due to their significant scientific interest. As we notify later, these galaxies are found to be less-massive ($\log_{10}{(M_*/M_\odot)}<10.3$) QGs found by using ground-based telescopes. 
The discussion based on spectroscopic observations regarding contamination, and luminosity functions are presented in \citet{Harikane24}.

\subsection{JWST/NIRSpec Spectroscopy}

This work uses JWST/NIRSpec data taken in a JWST program GO 1740 (PI: Y. Harikane). The NIRSpec spectra were taken in the fixed slit (FS) mode with S400A1 aperture, and the low-resolution ($R \sim 30 \text{--} 300$) mode with PRISM/CLEAR covering the wavelength range of $0.60 \text{--} 5.3 ~\si{\mu m}$. The spectra were obtained in the \texttt{2-POINT-NOD} mode with total exposure times are 2873 seconds for HD1, 1801 seconds for HD2, and 5041 seconds for HD3. These exposure times are adequate to detect the continuum and assess whether the galaxies are indeed LBGs or not.

We start from the level-2 data downloaded from \texttt{MAST}. We process the data using \texttt{jwst} pipeline version 1.14.0 with \texttt{the calibration reference data system (CRDS)} pipeline mapping context 1229.

For HD3, we have observed flux losses in the JWST/NIRSpec spectrum compared to the JWST/NIRCam photometry. We have found that the spatial width of the aperture used in the 1D spectrum extraction process is about half the length of the spatial extent of its spectrum. Following advice from the JWST help desk, we re-extract the 1D spectrum of HD3 by default box car extraction method with a larger aperture enough to capture whole spectrum. The spatial width of the aperture is defined based on a Gaussian fit to the spatial profile obtained by integrating the 2D spectrum over the wavelength direction.

\subsection{JWST/NIRCam and JWST/MIRI imaging}

\begin{figure*}
	\includegraphics[width=2\columnwidth]{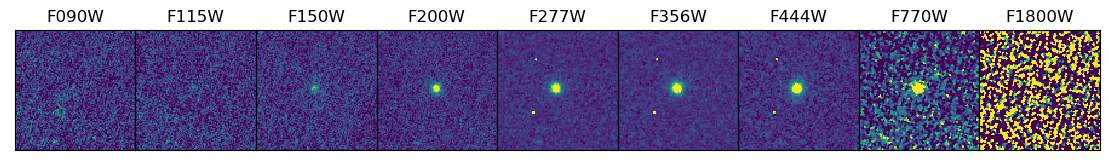}
    \caption{NIRCam and MIRI cutout images of HD3. Each image measures $4 \arcsec \times 4 \arcsec$. Two bright dots seen in F277W, F356W, F444W images are thought to be artifacts in images.}
    \label{fig:hd3_cutout}
\end{figure*}

Only for HD3, the NIRCam and MIRI images are available in the Public Release IMaging for Extragalactic Research (PRIMER, GO 1837, PI: J.S. Dunlop).
We use the mosaic images of F090W, F115W, F150W, F200W, F277W, F356W, and F444W filters of the NIRCam instrument, as well as F770W and F1800W filters of the MIRI instrument, which were downloaded from the Dawn JWST Archive (DJA)\footnote{\url{https://dawn-cph.github.io/dja/}} (mosaic version: v7.0).
In Figure~\ref{fig:hd3_cutout}, we show $4 \arcsec \times 4 \arcsec$ cutout images of HD3 for each available NIRCam and MIRI filters. 

\section{Analyses and Results}\label{sec:analysis}
\subsection{JWST/NIRSpec spectra for our targets}\label{sec:NIRSpec}
\begin{figure*}
    \begin{tabular}{c}
      \begin{minipage}[t]{1.2\columnwidth}
        \centering
        \includegraphics[width=\columnwidth]{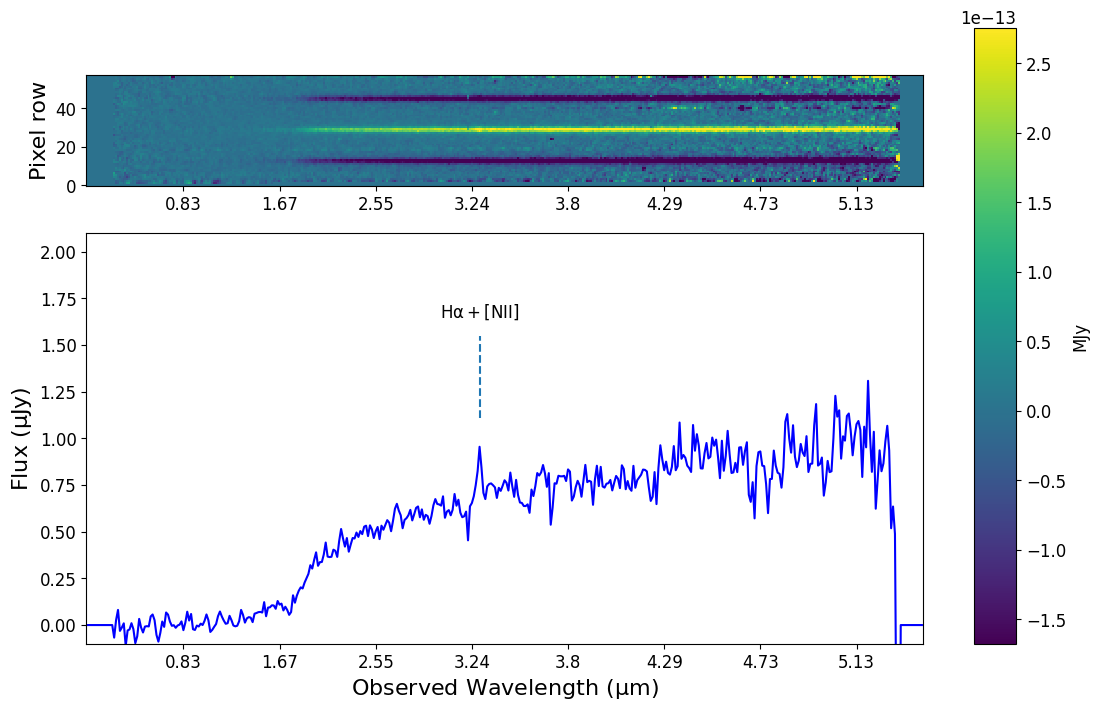}
        \subcaption{HD1}
      \end{minipage} \\
      \begin{minipage}[t]{1.2\columnwidth}
        \centering
        \includegraphics[width=\columnwidth]{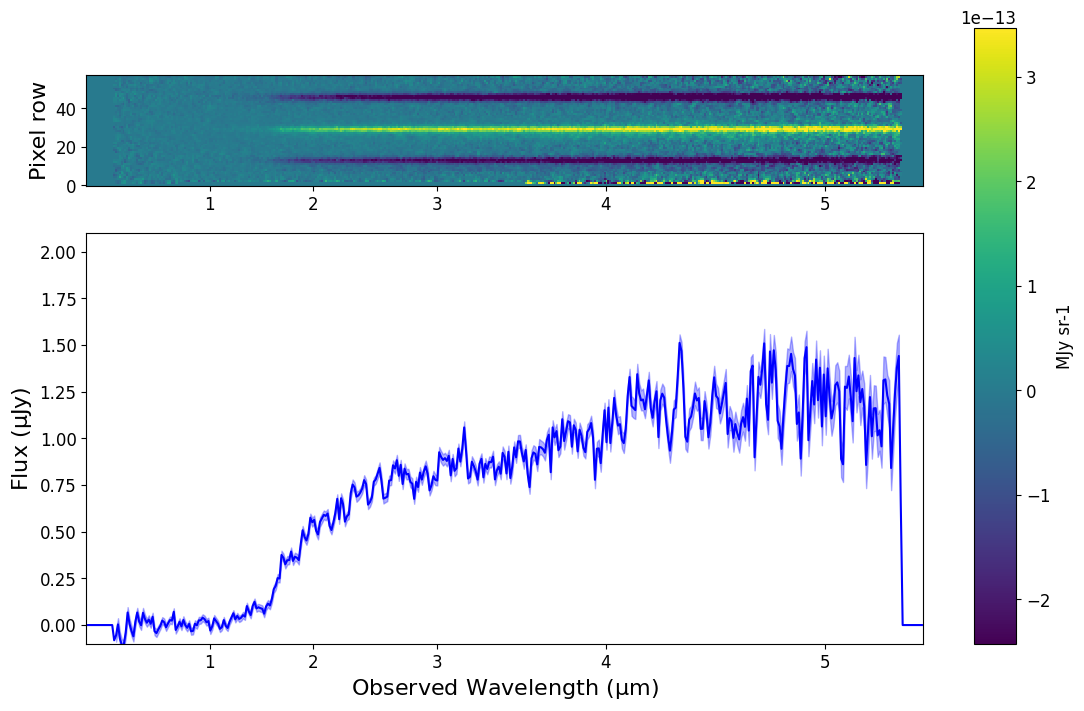}
        \subcaption{HD2}
      \end{minipage} \\
      \begin{minipage}[t]{1.2\columnwidth}
        \centering
        \includegraphics[width=\columnwidth]{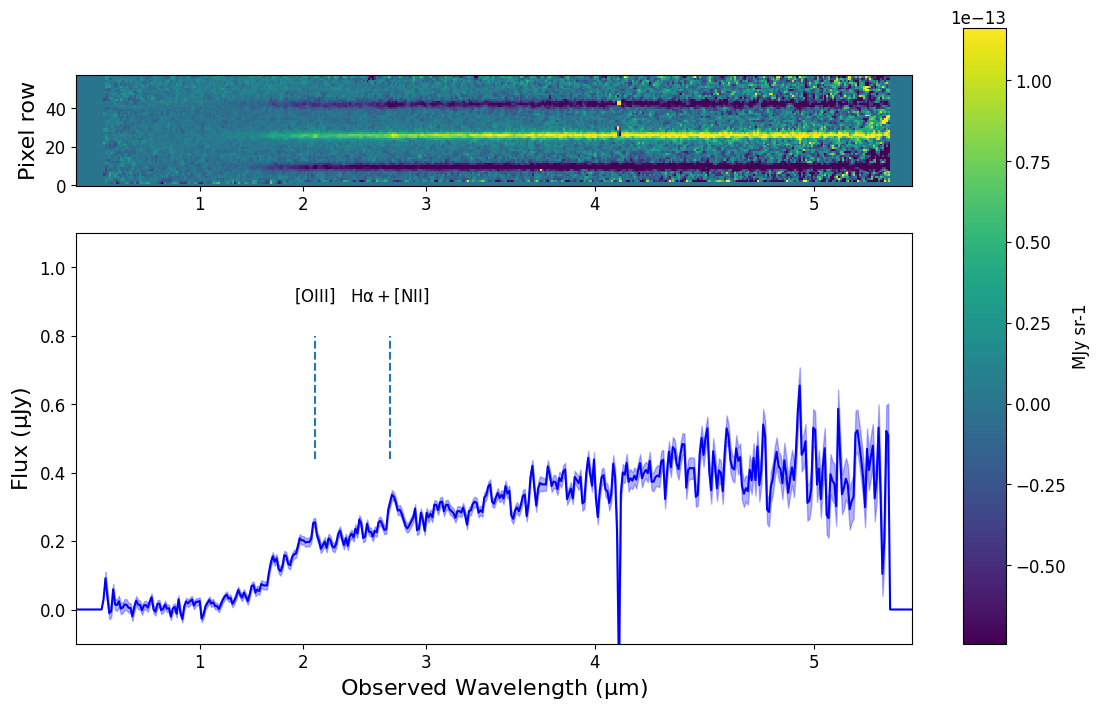}
        \subcaption{HD3}
      \end{minipage} 
    \end{tabular}
   \caption{1D and 2D spectra of the target galaxies, (a) HD1, (b) HD2, and (c) HD3. Each figure consists of two panels: the top panel displays the 2D spectrum, where the horizontal axis denotes the observed wavelength ($\si{\mu m}$), and the vertical axis illustrates the spatial position along the slit (pixel). The colour bar shows flux density in the unit of $\si{MJy}$ per $\si{sr}$. The bottom panel exhibits the 1D spectrum with the x-axis representing the observed wavelength ($\si{\mu m}$) and the y-axis representing the flux density ($\si{\mu Jy}$). The blue line illustrates the observed flux, and the blue shaded region surrounding it represents the 1-$\sigma$ error region.}
   \label{fig:spec}
\end{figure*}

 Our target spectra are presented in Figure~\ref{fig:spec} which shows that the continuum breaks of the galaxies in the short wavelength are not as sharp as those that would be expected for Lyman break. Therefore, the $H$-band dropout is actually due to a Balmer break. For HD1, we find an emission line feature at $3.2~\si{\mu m}$ in the NIRSpec spectrum (see Figure.~\ref{fig:spec}a). Given the redshift around 4 expected from the Balmer break wavelength, the emission line is likely to be blended H${\alpha}$+[NII], corresponding to the redshift $z=4.0$. In HD3 spectrum (Figure.~\ref{fig:spec}c), there is an emission line feature at $2.8~\si{\mu m}$. Given the redshift around 3 expected from the Balmer break, the emission line is likely to be blended H${\alpha}$+[NII], corresponding to the redshift $z=3.2$. Another feature is observed at $2.1~\si{\mu m}$, potentially corresponding to [OIII], for the case of $z=3.2$. We consider a negative signal at $4~\si{\mu m}$ to be caused by some error. We cannot find any significant line feature in HD2 spectrum. 

\subsection{JWST/NIRCam and JWST/MIRI photmetry of HD3}\label{sec:phot_hd3}
For HD3, we conduct aperture photometry by using photutils 1.5.0 \citep{photutils} in available images of NIRCam and MIRI using a $0.3 \arcsec$-radius aperture, larger than the effective diameter $2R_e$ in the F200W image obtained by using \texttt{Galfit}\footnote{\url{https://users.obs.carnegiescience.edu/peng/work/galfit/galfit.html}} \citep[][]{Peng02,Peng10} as described in \S\ref{sec:morHD3}. The aperture radius is larger than the PSF FWHM size of all images except for F1800W. The photometric values with $1\text{-}\sigma$ errors or $3\text{-}\sigma$ upper limits are shown in Table~\ref{tab:phot_jwst}. We estimate errors by performing aperture photometry randomly in the background region.

\renewcommand{\arraystretch}{1.5}
\begin{table*}
	\centering
	\caption{JWST/NIRCam and JWST/MIRI photometry of HD3. Flux densities or $3\text{-}\sigma$ upper limits are shown in the unit of nJy.}
	\label{tab:phot_jwst}
	\begin{tabular}{lcccccccccccc} 
		\hline
		ID & R.A. & Decl.  & \multicolumn{7}{c}{JWST/NIRCam} & & \multicolumn{2}{c}{JWST/MIRI}  \\
        \cline{4-10}  \cline{12-13} 
           &      &        & F090W & F115W & F150W & F200W & F277W & F356W & F444W & & F770W & F1800W \\
		\hline
		HD3 & 02:16:54.48 & $-$05:09:37.1 
        & $<55$ & $<44$ & $97 \pm 13$ & $220 \pm 13$  
        & $328 \pm 11$ & $405 \pm 10$ & $486 \pm 12$
        & & $536 \pm 42$ & $<457$\\
		\hline
	\end{tabular}
\end{table*}
\renewcommand{\arraystretch}{1.0}

\subsection{SED fitting}\label{sec:sed}

We conduct the spectral energy distribution (SED) fitting for the NIRSpec spectra. We employ the spectra spanning the range of $1.1$ to $5.0 ~\si{\mu m}$ in our analysis, removing both short and long wavelengths where the uncertainty is large. We apply 2-pixel binning for the spectra to match the actual resolution of the data. We use the SED fitting code \texttt{BAGPIPES}\footnote{\url{https://github.com/ACCarnall/bagpipes}} \cite[][]{Carnall18} to perform SED fitting and estimate various physical parameters of the galaxies. The prior distributions employed in the fitting process can be found in Table~\ref{tab:bgp_prior}. \texttt{BAGPIPES} relies on stellar population models from \citet{BC03}, particularly the 2016 version from \citet{BC16}, with IMF from \citet{Kroupa02}.
In our analysis, we incorporate the Calzetti model \citep{Calzetti00} to consider dust attenuation. Additionally, we adopt a delayed-$\tau$ star formation history (SFH) model with a form of the star formation rate (SFR) $\propto t\exp{(-t/\tau)}$. 

We perform SED fitting in two approaches. Given the limited spectral resolution ($R\sim100$) of our spectra and the fact that our targets are expected to be QGs without strong nebular emission lines, reliable fitting for metallicity and nebular parameters may be difficult. First, we perform the fitting with the metallicity value fixed at solar metallicity ($Z_*/Z_\odot = 1$), without nebular emission in our models. Secondary, we perform the fitting with the metallicity as a free parameter, still omitting nebular emission. In both cases, we conduct SED fitting with masking the line features.

For HD1, we set the redshift to $z=4.00$, determined from the position of the potential H${\alpha}$+[NII] emission line. For HD3, we fix the redshift at $z=3.20$ based on H${\alpha}$+[NII] emission line. However, for HD2, we define a redshift range from $z=2.0$ to $z=4.0$ based on the Balmer break position.

Table~\ref{tab:bgp_post} provides the values of the derived posterior distributions. Figures~\ref{fig:hd1_fit}, \ref{fig:hd2_fit}, and \ref{fig:hd3_fit} display the fitting outcomes for each of our targets.

\begin{table}
	\begin{center}
    	\caption{Free and pre-fixed physical parameters of spectral energy distribution (SED) fitting using \texttt{BAGPIPES}. The SED fitting is performed adopting a delayed tau star formation history (SFH) model and the Calzetti dust attenuation model.}
    	\label{tab:bgp_prior}
    	\begin{tabular}{lc} 
    		\hline
    		Parameter & Value Range \\
    		\hline
            Age$_{\text{del}}$ \ (Gyr)& $(0.01, 2.)$  \\
    		$\log_{10}{(M_*/M_\odot)}$ & $(8.,12.)$  \\
    		$\tau \ (\text{Gyr})$ & $(0.01, 2.)$ \\
    		$Z_*/Z_\odot$ & $1.$ (fixed) \\
            $A_V$ \ (mag) & $(0.,4.)$ \\
            $z$ & $(2.0,4.0) ^{a,b}$ \\
    		\hline
    	\end{tabular}\\
    \end{center}
    \footnotesize$^a$ For HD1, the redshift is fixed at $z=4.00$ estimated from H${\alpha}$ + [NII] emission line. \\
    \footnotesize$^b$ For HD3, the redshift is fixed at $z=3.20$ estimated from H${\alpha}$ + [NII] emission line. 
\end{table}

\renewcommand{\arraystretch}{1.5}
\begin{table*}
	\centering
	\caption{Physical properties of HD1, HD2, and HD3 obtained from SED fitting. $\text{SFR}$ is the star formation rate and $\text{sSFR}$ is specific star formation rate at the observed redshift. Values of SFR or $3\text{-}\sigma$ upper limit of SFR are shown in this table. The $\text{age}$ is defined as mass-weighted age. $t_\text{form} $ and $ t_\text{quench}$ are defined as the age of the Universe corresponding to the time of formation and quenching, respectively.
}
	\label{tab:bgp_post}
	\begin{tabular}{lccccccccc} 
		\hline

        ID
        & $\text{Age}$ \ (Gyr) 
        & $\log_{10}{(M_*/M_\odot)}$ 
        & $\text{SFR}/M_\odot ~\text{yr}^{-1}$       
        & $\log_{10}{(\text{sSFR}/\text{yr}^{-1})}$  
        & $Z/Z_\odot$  
        & $A_V$ \ (mag) 
        & $t_\text{quench}$ \ (Gyr)
        & $t_\text{form}$ \ (Gyr)
        & $z$\\
		\hline
        \multicolumn{10}{l}{Fixed metallicity case}\\
		\hline
  
        HD1 
        & ${0.72}${\raisebox{0.5ex}{\tiny$\substack{+0.05 \\ -0.03}$}}     
        & ${10.20}${\raisebox{0.5ex}{\tiny$\substack{+0.00 \\ -0.01}$}}     
        & ${<0.12}$
        & ${-12.59}${\raisebox{0.5ex}{\tiny$\substack{+1.05 \\ -3.02}$}}     
        & $1$ 
        & ${0.25}${\raisebox{0.5ex}{\tiny$\substack{+0.04 \\ -0.04}$}}   
        & ${1.07}${\raisebox{0.5ex}{\tiny$\substack{+0.07 \\ -0.09}$}}        
        & ${0.79}${\raisebox{0.5ex}{\tiny$\substack{+0.03 \\ -0.05}$}}     
        & ${4.00}$  \\

		HD2      
        & ${0.61}${\raisebox{0.5ex}{\tiny$\substack{+0.03 \\ -0.04}$}}        
        & ${10.16}${\raisebox{0.5ex}{\tiny$\substack{+0.00 \\ -0.00}$}}       
        & ${0.00}$
        & ${-15.72}${\raisebox{0.5ex}{\tiny$\substack{+2.76 \\ -5.14}$}}   
        & $1$ 
        & ${0.44}${\raisebox{0.5ex}{\tiny$\substack{+0.04 \\ -0.02}$}}        
        & ${1.47}${\raisebox{0.5ex}{\tiny$\substack{+0.07 \\ -0.07}$}}  
        & ${1.29}${\raisebox{0.5ex}{\tiny$\substack{+0.04 \\ -0.03}$}}     
        & ${3.29}${\raisebox{0.5ex}{\tiny$\substack{+0.01 \\ -0.01}$}} [1]  \\

        HD3 
        & ${0.55}${\raisebox{0.5ex}{\tiny$\substack{+0.11 \\ -0.09}$}}      
        & ${9.70}${\raisebox{0.5ex}{\tiny$\substack{+0.05 \\ -0.05}$}}  
        & ${0.36}${\raisebox{0.5ex}{\tiny$\substack{+0.39 \\ -0.29}$}}       
        & ${-10.16}${\raisebox{0.5ex}{\tiny$\substack{+0.30 \\ -0.71}$}}     
        & $1$ 
        & ${0.35}${\raisebox{0.5ex}{\tiny$\substack{+0.25 \\ -0.23}$}}      
        & ${1.89}${\raisebox{0.5ex}{\tiny$\substack{\\ -0.14}$}}      
        & ${0.78}${\raisebox{0.5ex}{\tiny$\substack{+0.09 \\ -0.11}$}}       
        & ${3.20}$  \\
        
		\hline
       \multicolumn{10}{l}{Free metallicity case}\\
       \hline

        HD1 
        & ${0.63}${\raisebox{0.5ex}{\tiny$\substack{+0.03 \\ -0.03}$}}     
        & ${10.23}${\raisebox{0.5ex}{\tiny$\substack{+0.02 \\ -0.02}$}}     
        & ${0.00}$   
        & ${-17.82}${\raisebox{0.5ex}{\tiny$\substack{+4.03 \\ -6.76}$}}     
        & ${0.08}${\raisebox{0.5ex}{\tiny$\substack{+0.03 \\ -0.03}$}} 
        & ${0.83}${\raisebox{0.5ex}{\tiny$\substack{+0.07 \\ -0.07}$}}   
        & ${1.02}${\raisebox{0.5ex}{\tiny$\substack{+0.08 \\ -0.06}$}}        
        & ${0.88}${\raisebox{0.5ex}{\tiny$\substack{+0.03 \\ -0.03}$}}     
        & ${4.00}$  \\
        
		HD2      
        & ${0.82}${\raisebox{0.5ex}{\tiny$\substack{+0.05 \\ -0.06}$}}        
        & ${10.08}${\raisebox{0.5ex}{\tiny$\substack{+0.01 \\ -0.00}$}}       
        & ${<0.03}$
        & ${-18.02}${\raisebox{0.5ex}{\tiny$\substack{+4.35 \\ -9.88}$}}   
        & ${2.05}${\raisebox{0.5ex}{\tiny$\substack{+0.05 \\ -0.05}$}}  
        & ${0.01}${\raisebox{0.5ex}{\tiny$\substack{+0.01 \\ -0.00}$}}        
        & ${1.35}${\raisebox{0.5ex}{\tiny$\substack{+0.10 \\ -0.08}$}}  
        & ${1.17}${\raisebox{0.5ex}{\tiny$\substack{+0.04 \\ -0.02}$}}     
        & ${3.17}${\raisebox{0.5ex}{\tiny$\substack{+0.01 \\ -0.01}$}} [1]  \\
        
		HD3      
        & ${0.68}${\raisebox{0.5ex}{\tiny$\substack{+0.09 \\ -0.14}$}}        
        & ${9.66}${\raisebox{0.5ex}{\tiny$\substack{+0.05 \\ -0.04}$}}       
        & ${0.05}${\raisebox{0.5ex}{\tiny$\substack{+0.31 \\-0.05}$}}        
        & ${-10.95}${\raisebox{0.5ex}{\tiny$\substack{+0.82 \\ -2.43}$}}   
        & ${0.21}${\raisebox{0.5ex}{\tiny$\substack{+0.03 \\ -0.02}$}}  
        & ${0.40}${\raisebox{0.5ex}{\tiny$\substack{+0.25 \\ -0.20}$}}        
        & ${1.72}${\raisebox{0.5ex}{\tiny$\substack{+0.17 \\ -0.20}$}}  
        & ${1.29}${\raisebox{0.5ex}{\tiny$\substack{+0.14 \\ -0.02}$}} 
        & ${3.20}$  \\
		\hline
	\end{tabular}
    \footnotesize{[1] Systematic errors are adopted from \citet{Glazebrook24}.} \\
\end{table*}
\renewcommand{\arraystretch}{1.0}

\subsubsection{HD1}\label{sec:hd1}

\begin{figure*}
	\includegraphics[width=1.5\columnwidth, angle=270]{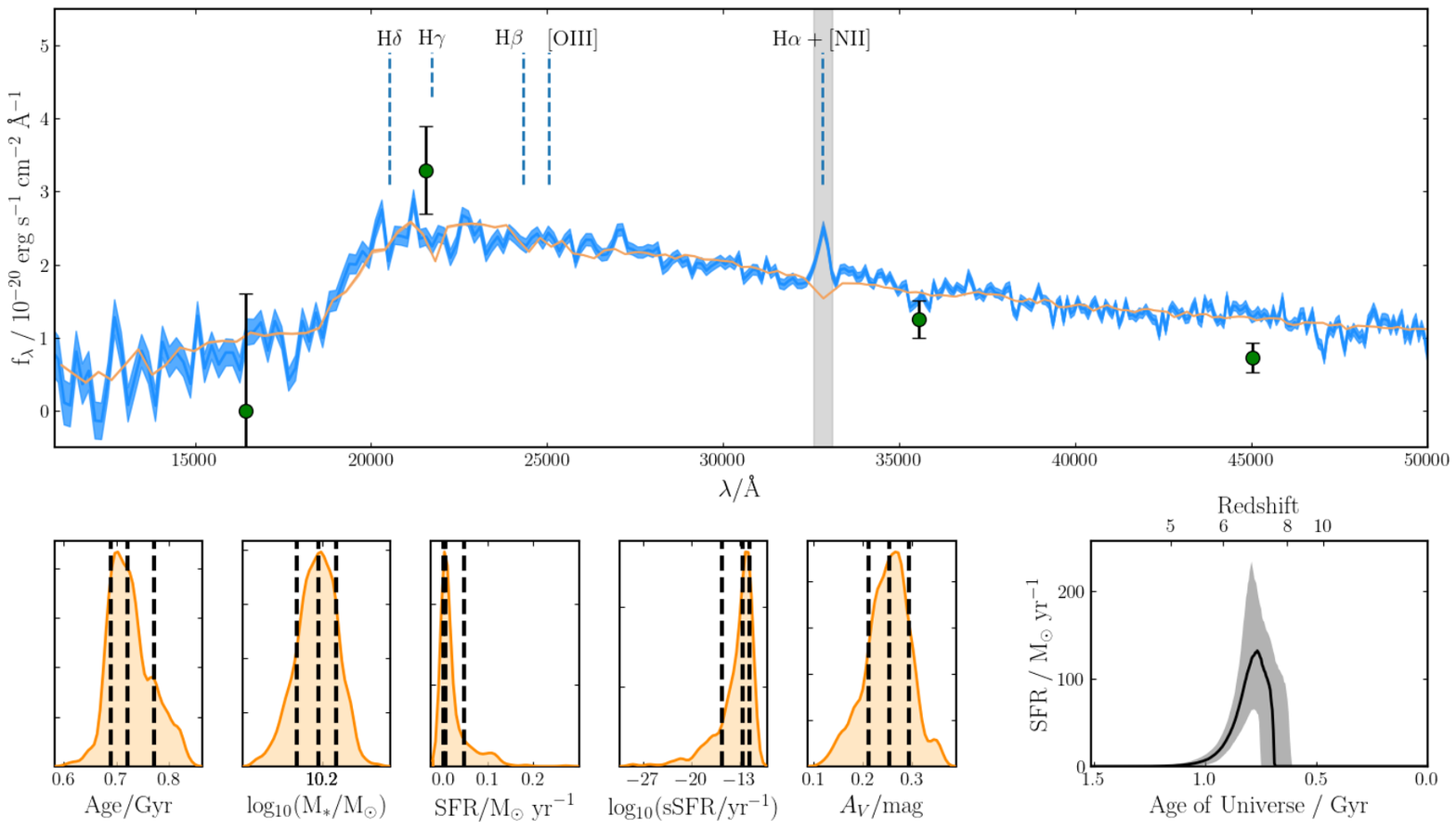}
    \caption{Fit result for HD1. In the top figure, the blue line corresponds to the observed spectrum data without binning, whereas the orange line represents the best-fit SED, and the green plots represent photometric data points with 1-$\sigma$ error bars from \citet{Harikane22a}, especially VISTA/$H,K_s$-band, and Spitzer/IRAC [3.6] and [4.5]. Lightgrey shaded region represents a masked region which is not used in this fitting. The bottom-left five panels show the probability distributions of the parameters. The dashed lines correspond to the posterior percentiles of 16th, 50th, and 84th. The bottom-right panel illustrates the star formation history (SFH), featuring the median posterior of the best-fit SFH depicted by the black line. The grey shaded region represents the 14th–68th percentile posterior distribution from the best-fit SFH. }
    \label{fig:hd1_fit}
\end{figure*}

As mentioned in \S\ref{sec:NIRSpec} (see also Figure~\ref{fig:spec}), the spectrum of HD1 exhibits a notable emission line identified as H${\alpha}$+[NII] emission. In order to minimise the influence of this emission line on the fitting process, we conduct SED fitting with masking the line as shown by the grey shade in the top panel of Figure~\ref{fig:hd1_fit}. 
Another point to note is that the photometric data points are not perfectly matched to the spectrum as found in the top panel of Figure~\ref{fig:hd1_fit}. Indeed, the photometric data of $K_s$, [3.6] and [4.5] shows bluer color than the spectrum. 
This discrepancy might be attributed to any systematic uncertainty of previous lower resolution and signal-to-noise ratio photometric data. In this paper, we consider NIRSpec data to be more reliable than the photometric data.

As summarised in Table~\ref{tab:bgp_post} and Figure~\ref{fig:hd1_fit}, the stellar mass of HD1 is $\log{(M_*/M_\odot)}={10.20}${\raisebox{0.5ex}{\tiny$\substack{+0.00 \\ -0.01}$}}, a relatively modest value when compared to QGs typically observed with ground-based telescopes. 
The 3-sigma upper limit of star formation rate (SFR) of HD1 is $\text{SFR}< {0.12}~M_\odot ~\text{yr}^{-1}$. 
As a result, we find a specific star formation rate (sSFR) of $\log{(\text{sSFR}/\text{yr}^{-1})}={-12.59}${\raisebox{0.5ex}{\tiny$\substack{+1.05 \\ -3.02}$}}.
This is consistent with a quiescent scenario because the sSFR value is more than 2-dex lower than that expected for the star-forming main-sequence (SFMS) at redshift $z=4$ as shown in Figure~\ref{fig:sfms}. 
Other spectroscopically confirmed QGs at $z\simeq4$ tend to be more massive, with $\log{(M_*/M_\odot)}\gtrsim{10.3}$ \citep[][]{Tanaka19, Valentino20, Carnall23b,Antwi-Danso23, Kakimoto24}. HD1 is the least massive QG observed at $z\simeq4$ so far.
We list the time of formation $t_\text{form}$ and the time of quenching $t_\text{quench}$ in Table~\ref{tab:bgp_post}, adopting the definitions by \citet{Carnall18}. The time of formation $t_\text{form}$ is defined as
\begin{equation}
t_\text{form} = t(z_\text{form}) = \frac{\int^{t_\text{obs}}_0 t SFR(t) dt}{\int^{t_\text{obs}}_0 SFR(t) dt},
\end{equation}
where $t_\text{obs}=t(z_\text{obs})$ is the age of universe at the redshift of observation $z_\text{obs}$. $z_\text{form}$ corresponds to the redshift at which the stellar mass of the galaxy would have been half of its present mass.
The time of quenching $t_\text{quench}$ is the age of universe corresponding to the time where a normalised $\text{SFR}$ becomes less than $10\%$ of averaged $\text{SFR}$ across the history of the galaxy.
The timescale for quenching is defined as $\Delta t_\text{quench}=t_\text{quench}-t_\text{form}$ \citep{Carnall18}. 
It is approximately $0.3~\si{Gyr}$. The mass-weighted age of HD1 is $0.72~\si{Gyr}$.
Corresponding redshifts for the formation and quenching times are $z_\text{form} = {6.7}$, and $z_\text{quench}= {5.3}$, respectively. 

In the fitting with the free metallicity parameter, we also obtain a quiescent solution with a stellar mass of  $\log{(M_*/M_\odot)}={10.23}${\raisebox{0.5ex}{\tiny$\substack{+0.02 \\ -0.02}$}}. The stellar mass and the SFR is almost same as those with the solar metallicity presented above. The metallicity is estimated as ${0.08}${\raisebox{0.5ex}{\tiny$\substack{+0.03\\ -0.03}$}}$~Z_\odot$. Dust attenuation $A_V$ is estimated to be higher ($A_V=0.83~\text{mag}$) compared to when the metallicity is fixed at the solar value ($A_V=0.25~\text{mag}$). This is because a bluer colour expected from the sub-solar metallicity needs to be compensated by the higher dust attenuation. The mass-weighted age is slightly younger with a value of $0.63~\si{Gyr}$.

\subsubsection{HD2}

\begin{figure*}
	\includegraphics[width=1.5\columnwidth, angle=270]{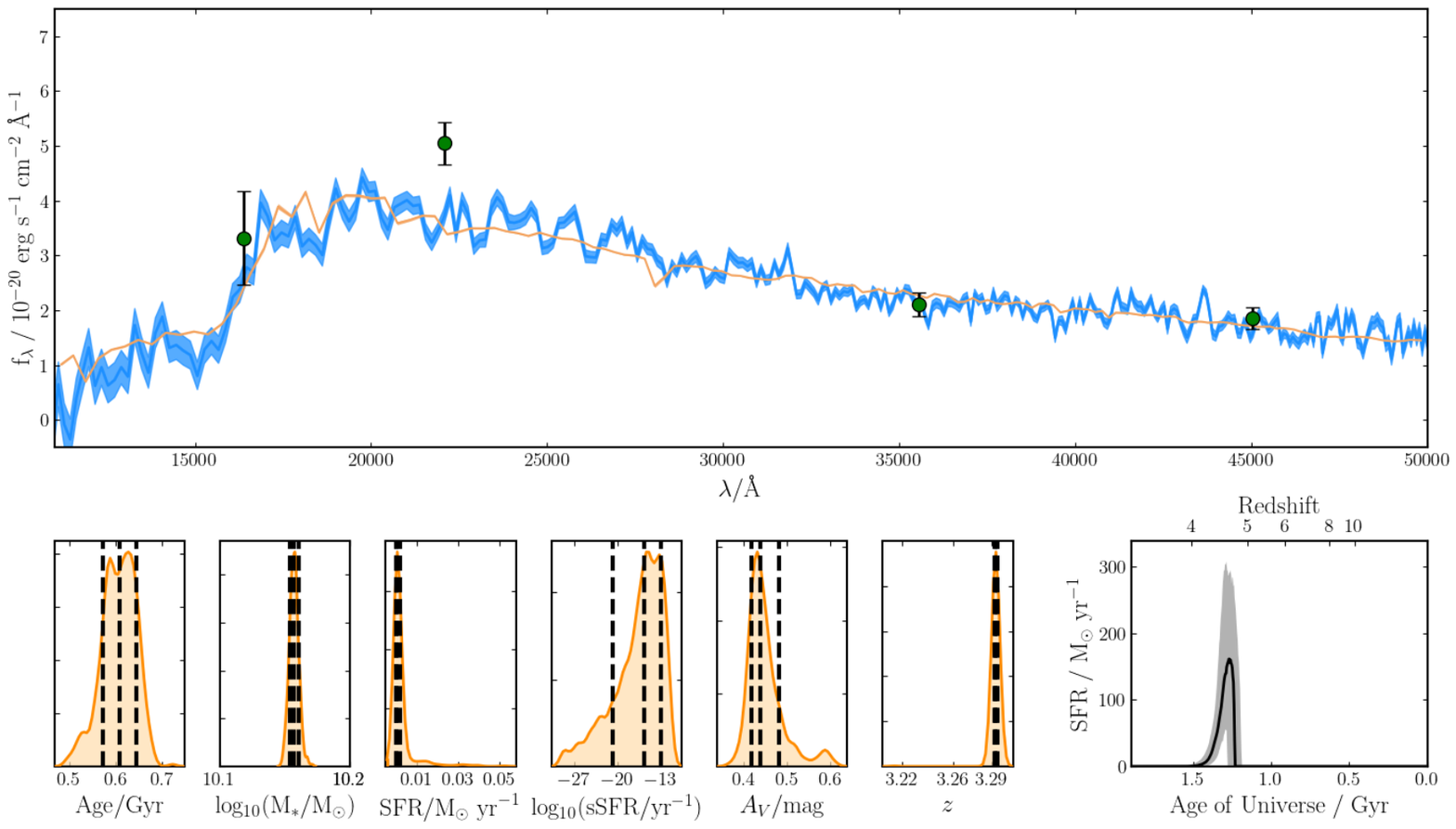}
    \caption{Same as Figure~\ref{fig:hd1_fit} but for HD2, where the redshift is another free parameter. $H,K$-band data are from UKIRT.}
    \label{fig:hd2_fit}
\end{figure*}

In Figure~\ref{fig:hd2_fit}, on the contrary to HD1, the spectrum and photometry seem to match, except for the $K$ band. Since the $K$ band photometric point is brighter, the photometric colour between $K$, [3.6], and [4.5] is bluer than the spectrum, mimicking an LBG solution.
The stellar mass of HD2 is $\log{(M_*/M_\odot)}={10.16}${\raisebox{0.5ex}{\tiny$\substack{+0.00 \\ -0.00}$}} with a redshift of $z=3.29$. 
HD2 exhibits the most significant quiescence because of the $\text{SFR}= {0.00}~M_\odot ~\text{yr}^{-1}$.
We find $\log{(\text{sSFR}/\text{yr}^{-1})}={-15.72}${\raisebox{0.5ex}{\tiny$\substack{+2.26 \\ -5.14}$}}, lowest value among our samples.
This value is more than 2-dex lower than SFMS at $z=3$ as shown in Figure~\ref{fig:sfms} and consistent with a quiescent scenario. 
The mass-weighted age of HD2 is $0.61~\si{Gyr}$. The formation and quenching times listed in Table~\ref{tab:bgp_post} correspond to the formation and quenching redshifts of $z_\text{form} =4.6$ and $z_\text{quench} = 4.1$, respectively. The quenching timescale $\Delta t_\text{quench}$ is $0.2~\si{Gyr}$.

In the fitting with the free metallicity parameter, we also obtain a quiescent solution with a stellar mass of  $\log{(M_*/M_\odot)}={10.08}${\raisebox{0.5ex}{\tiny$\substack{+0.01 \\ -0.00}$}}. The stellar mass and the SFR is almost same as the solar metallcity case. Redshift is estimated as $z=3.17$. The metallicity is estimated as ${2.05}${\raisebox{0.5ex}{\tiny$\substack{+0.05\\ -0.05}$}}$~Z_\odot$. Dust attenuation $A_V$ is estimated to be lower ($A_V=0.01~\text{mag}$) compared to when the metallicity is fixed at the solar value ($A_V=0.44~\text{mag}$). An intrinsic redder colour expected from the higher metallicity needs less dust attenuation. The mass-weighted age is older with a value of $0.82~\si{Gyr}$.

\subsubsection{HD3}\label{sec:hd3}

\begin{figure*}
	\includegraphics[width=1.5\columnwidth, angle=270]{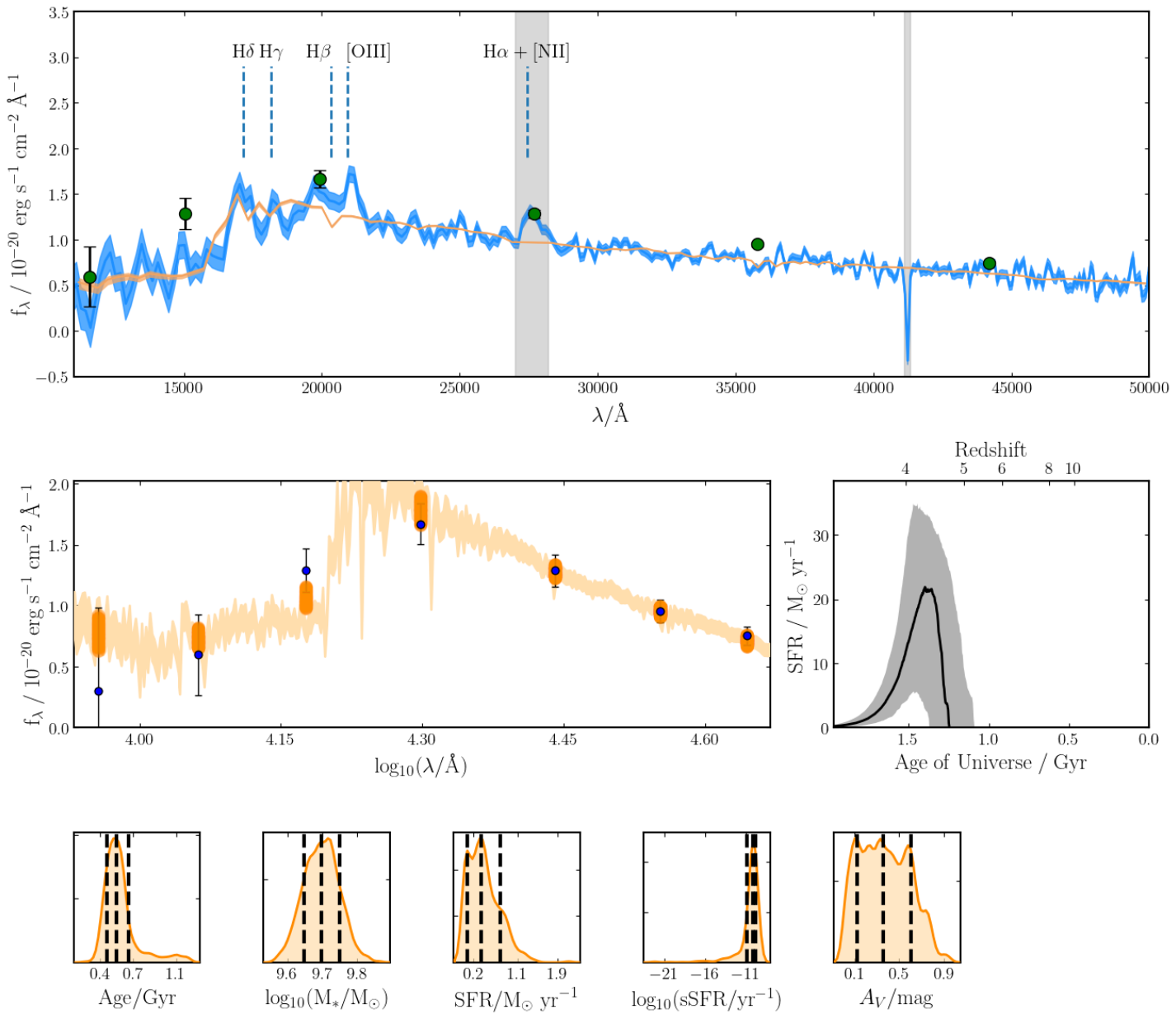}
    \caption{Same as Figure~\ref{fig:hd1_fit} but for HD3. The photometric data in the top panel are our measurements based on NIRCam images (Table~\ref{tab:phot_jwst}). The orange line is the best-fit SED without calibration correction (see text). The middle left figure indicates the 16th--84th percentile posterior distribution from the best-fit SED.}
    \label{fig:hd3_fit}
\end{figure*}

\begin{figure*}
	\includegraphics[width=1\columnwidth, angle=270]{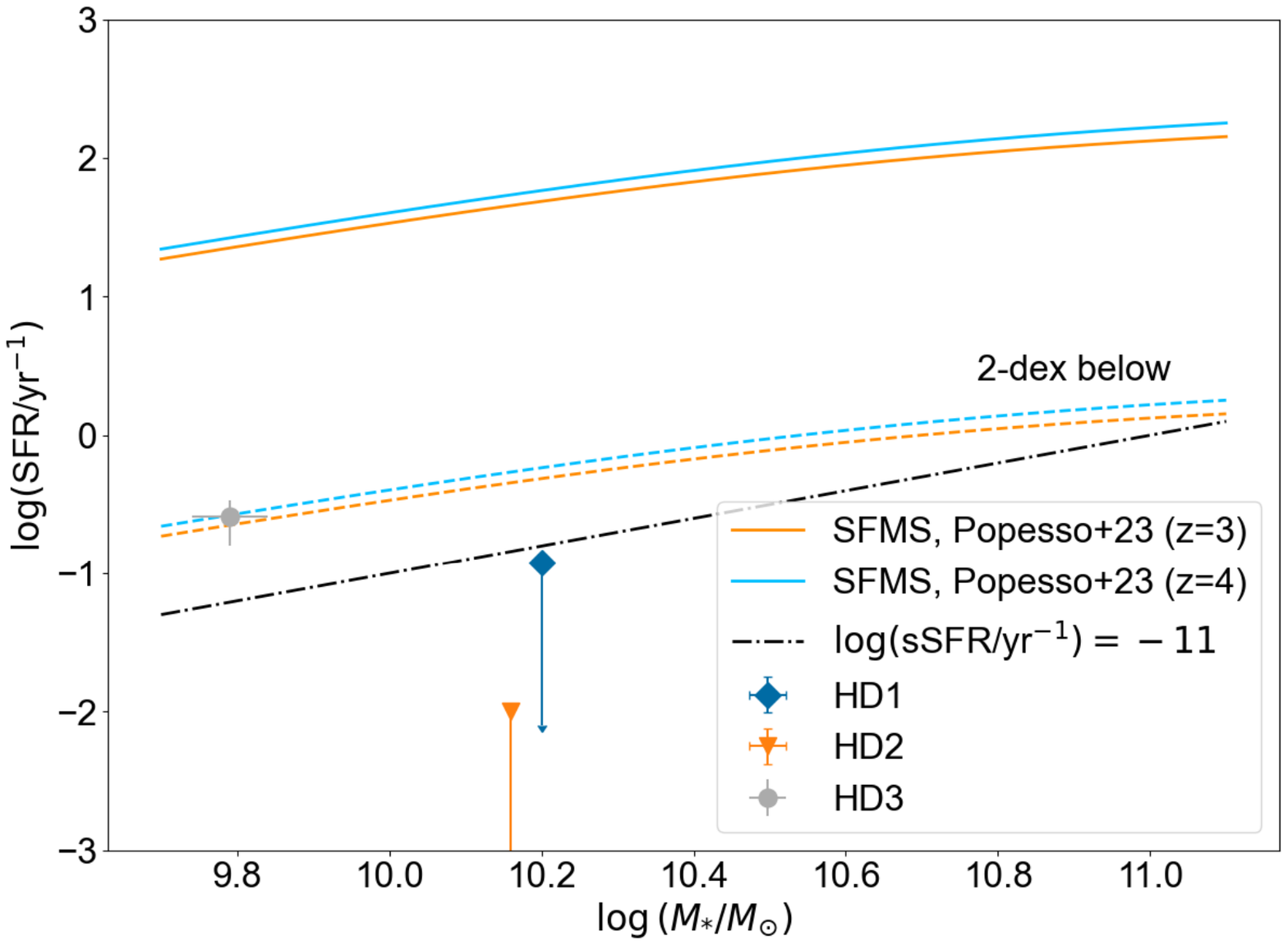}
    \caption{The figure showing SFR versus $M_{*}$ of the target galaxies. The orange and blue lines represent the star-forming main sequence (SFMS) at redshifts $z=3$ and $z=4$, , respectively, as derived from equation (14) in \citet{Popesso23}. Dashed lines indicate 2-dex below the SFMS, with each color corresponding to the respective redshifts. The black dashed dotted line denotes the threshold at $\log{(\text{sSFR}/\text{yr}^{-1})}={-11}$.}
    \label{fig:sfms}
\end{figure*}

Rather than directly correcting the observed spectra using photometric data, we apply a photometric calibration to HD3 by including second-order Chebyshev polynomial terms in our fitting. In order to minimise the influence of this emission line on the fitting process, we conduct SED fitting with masking the line and artifact as shown by the grey shade in the top panel of Figure~\ref{fig:hd3_fit}. This calibration is performed considering only the available NIRCam photometry for HD3. As shown in the left middle panel of Figure~\ref{fig:hd3_fit}, the NIRCam photometric data and the model spectrum are in good agreement with each other after systematic correction by the second-order Chebyshev polynomial function in {\tt BAGPIPES}, except for F150W which is also consistent within a $\sim2\sigma$ range.
Based on the information provided in Table~\ref{tab:bgp_post} and Figure~\ref{fig:hd3_fit}, it is evident that HD3 stands out as the least massive galaxy (i.e. $M_*<{10}^{10}M_\odot$) among the target galaxies with the stellar mass of $\log{(M_*/M_\odot)}={9.70}${\raisebox{0.5ex}{\tiny$\substack{+0.05 \\ -0.05}$}}. HD3 has an $\text{SFR} = {0.36}${\raisebox{0.5ex}{\tiny$\substack{+0.39 \\ -0.29}$}}$~M_\odot ~\text{yr}^{-1}$ and an sSFR of $\log{(\text{sSFR}/\text{yr}^{-1})}={-10.16}${\raisebox{0.5ex}{\tiny$\substack{+0.30 \\ -0.71}$}}. This value is more than 2-dex lower than SFMS at $z=3$ as shown in Figure~\ref{fig:sfms} and consistent with a quiescent/post-starburst scenario. 
The mass-weighted age of HD3 is $0.55~\si{Gyr}$. The formation and quenching times in Table~\ref{tab:bgp_post} correspond to the formation and quenching redshifts of $z_\text{form} =4.2$ and $z_\text{quench} =3.3$, respectively. The quenching timescale $\Delta t_\text{quench}$ is $0.4~\si{Gyr}$.

In the fitting with the free metallicity parameter, we also obtain a quiescent solution with a stellar mass of  $\log{(M_*/M_\odot)}={9.66}${\raisebox{0.5ex}{\tiny$\substack{+0.05 \\ -0.04}$}}. The stellar mass and the SFR is almost same as those obtained with the solar metallicity. The metallicity is estimated as ${0.21}${\raisebox{0.5ex}{\tiny$\substack{+0.03\\ -0.02}$}}$~Z_\odot$. Dust attenuation $A_V$ is estimated to be slightly higher ($A_V=0.40~\text{mag}$) compared to the solar metallicity case ($A_V=0.35~\text{mag}$) to compensate a bluer intrinsic colour from the sub-solar metallicity. The mass-weighted age is older with a value of $0.68~\si{Gyr}$.

\subsection{Emission line features of HD1 and HD3}\label{sec:ha}

Utilising \texttt{specutils} \citep{spectutils}, we quantify the line flux and the rest-frame equivalent width (EW) of the emission line features of HD1 and HD3 as summarised in Table~\ref{tab:hd1_ha}.

We fit the line features by a single Gaussian profile to estimate the full-width-at-half-maximum (FWHM). The line fluxes are calculated by integrating the flux at the peak of the line feature over a width of $2$FWHM. EWs are also estimated over the same range as the line fluxes. The uncertainties are estimated as the standard deviations of the distributions of fluxes or EWs repeatedly measured at some random wavelength positions in each spectrum.

We find an emission line that seems to be the H${\alpha}$+[NII] emission line in the HD1 spectrum (see Figure~\ref{fig:hd1_fit}). We fit the line and have an FWHM of $2600~\si{km.s^{-1}}$, which is almost as same as the spectral resolution at the observed wavelength, indicating that the line is not resolved in this observation. The wavelength integration width is always set as $5~\si{pix}$, which is approximately twice the FWHM for this line.

We also find emission line features that seem to be the H${\alpha}$+[NII] and H${\beta}$+[OIII] emission line in the HD3 spectrum (see Figure~\ref{fig:hd3_fit}). For H${\alpha}$+[NII] emission line feature, we fit the line and have an FWHM of $6600~\si{km.s^{-1}}$. The spectral resolution at the observed wavelength is $3600~\si{km.s^{-1}}$ indicating that the line is resolved. The wavelength integration width is always set as $8~\si{pix}$, which is approximately twice the FWHM for this line. For the [OIII] emission line feature, we consider it to be the [OIII]$\lambda \lambda 4959, 5007$ lines based on the position of H$\beta$, which is out of range from this feature. For this feature, we fit the line and have an FWHM of $3600~\si{km.s^{-1}}$. The spectral resolution at the observed wavelength is  $5700~\si{km.s^{-1}}$  indicating that the line remains unresolved. The wavelength integration width is always set as $4~\si{pix}$, which is approximately twice the FWHM for this line. 

In our spectra of HD1 and HD3, we plot the lines at the expected positions of H$\beta$, H$\gamma$, and H$\delta$ for reference. However, these lines have low significance in our spectra, so we do not analyse them in this paper.

\begin{table}
	\centering
	\caption{Line flux and rest-frame equivalent width of emission line features in the HD1 and HD3 spectra.}
	\label{tab:hd1_ha}
	\begin{tabular}{cccc} 
		\hline

        ID
        &
        Line
        & $\text{Flux}~(10^{-20}~\si{erg.cm^{-2}.s^{-1}})$
        & $\text{EW}_0~(\si{\angstrom})$ \\
		\hline 

        HD1
        &
        \text{H${\alpha}$+[NII]}
        & $44 \pm 8$      
        & $24 \pm 4$  \\

        HD3
        &
        \text{H${\alpha}$+[NII]}
        & $59 \pm 11$      
        & $61 \pm 12$  \\
        
        &
        \text{[OIII]}
        & $47 \pm 6$      
        & $37 \pm 6$  \\
        
		\hline
	\end{tabular}
\end{table}

\subsection{Morphology of HD3}\label{sec:morHD3}

Since we have high spatial resolution NIRCam images of HD3, we examine the morphological features of HD3 using \texttt{Galfit} \citep[][]{Peng02,Peng10}, a 2D profile fitting code. In this fitting procedure, we use a point spread function (PSF) generated by stacking of stars in the same image. 
Generally, the galaxy size depends on the observed wavelength at $z<3$ \citep[e.g.][]{vanderWel14}  To ensure consistency with the rest-frame wavelength in \citet{vanderWel14}, we assessed the morphology in the F150W, F200W, F277W, F356W images. We fit HD3 with a single S\'{e}rsic profile with free parameters: position, S\'{e}rsic index, effective radius, total magnitude, axis ratio, and position angle. The resulting morphological parameters for the images are outlined in Table~\ref{tab:hd3_morph}.
The effective radius $R_{e,{\text{F200W}}}$ is approximately $0.96~\si{kpc}\sim 0.13~\si{arcsec}$ and is sufficiently larger than empirical PSF FWHM size ($0.1~\si{arcsec}$), meaning HD3 is resolved. S\'{e}rsic index is estimated to be $n \simeq 1$, suggesting that HD3 shows a disc-like morphology.

Interestingly, the effective radius becomes smaller in longer wavelength. This declining trend of size regarding to the wavelength seen in some studies (e.g. \citet{vanderWel14, Ito23}. Moreover, S\'{e}rsic index becomes larger at longer wavelength. These trends might be related to spatial distribution of past star formation within the galaxy. Since longer wavelengths trace older stellar populations, more centrally concentrated brightness profiles in longer wavelengths suggest such a distribution of older stellar populations, indicating the inside-out formation process \citep[e.g.][]{Nelson12,Nelson16}.

\begin{table}
	\centering
	\caption{ Morphological parameters of HD3 obtained from Galfit. $R_e$, $n$, and $q$ denote the derived effective radius, S\'{e}rsic index, and axis ratio, respectively.} 
	\label{tab:hd3_morph}
	\begin{tabular}{cccc} 
		\hline

        Filter
        &$R_e$ \ (kpc)
        & $n$ \\
		\hline 

        F150W 
        &$0.948 \pm 0.102$      
        & $1.09 \pm 0.18$ \\
        
        F200W 
        &$0.959 \pm 0.036$      
        & $1.07 \pm 0.07$ \\
        
        F277W 
        &$0.797 \pm 0.021$      
        & $1.06 \pm 0.11$ \\

        F356W 
        &$0.687 \pm 0.027$      
        &$1.63 \pm 0.20$ \\
        
        F444W 
        &$0.648 \pm 0.021$      
        &$1.79 \pm 0.17$ \\
        
		\hline
	\end{tabular}
\end{table}

\section{Discussion}\label{sec:discussion}

\subsection{\texorpdfstring{Origin of emission lines in HD1 and HD3 spectra}{Origin of emission line sin HD1 and HD3 spectra}}\label{sec:origin}

In this section, we discuss possible origin of the H${\alpha}$+[NII] emission line feature seen in HD1 and HD3 spectra. An important thing to note is that we cannot estimate [NII]/H$\alpha$ from our data, but there are two possibilities that this line is H$\alpha$ dominant or [NII] dominant.

Firstly, we discuss possibilities of H$\alpha$ dominant case. In this case, there are two possibilities that the H$\alpha$ line is from AGN or star formation. 

QGs exhibiting H$\alpha$ emission lines were identified by \citet{Carnall23b, Nanayakkara24, Carnall24, deGraaff24}.  In \citet{Nanayakkara24}, one QG displays prominent emission lines, suggesting the possibility of AGN activity. Moreover, seven other QGs show [OIII]$\lambda5007$ or H$\beta$ emission lines with $S/N>3$, and four QGs show blended H${\alpha}$+[NII] emission lines. In \citet{Carnall23b}, H$\alpha$ and [NII] emission lines, as well as a marginal [OIII] emission line, were observed in a QG at $z=4.658$, with a clear broad H$\alpha$ component. The comparison of the strength of the H$\alpha$ narrow component with [NII] also indicates the presence of an AGN. Based on these findings, we consider the possibility of AGNs existing in our samples.

Assuming that most of the H$\alpha$ flux comes from a broad line region of the AGN, we can estimate the black hole mass using following formula in \citet{GreeneHo05},
\begin{multline}
M_{\text{BH}} = 2.0 \times {10}^{6} \\ \times \left(\frac{L_{\text{H}\alpha}}{{10}^{42}~\si{erg.s^{-1}}}\right)^{0.55}\left(\frac{\text{FWHM}_{\text{H}\alpha}}{{10}^{3}~\si{km.s^{-1}}}\right)^{2.06}~M_\odot.
\end{multline}
For HD1, we have an upper limit value of $\log{(M_\text{BH}/M_\odot)} \simeq 6$ where FWHM of H${\alpha}$+[NII] line is estimated as $2800~\si{km.s^{-1}}$ from single Gaussian fitting. This FWHM is almost same as the wavelength resolution in the spectrum around this wavelength range $\simeq 2500~\si{km.s^{-1}}$, making our estimation an upper limit of the $M_{\text{BH}}$. 
Some studies \citep{deGraaff24, Glazebrook24, Nanayakkara24} report that the spectral resolution of NIRSpec observation for objects that are much smaller than the slit width, can be higher than the values presented by STScI \footnote{\url{https://jwst-docs.stsci.edu/jwst-near-infrared-spectrograph/nirspec-instrumentation/nirspec-dispersers-and-filters}}. Although, we can not measure the size of HD1, but if this is applicable to our targets, this emission might be resolved.
For HD3, we have an upper limit value of $\log{(M_\text{BH}/M_\odot)} \simeq 7$, where FWHM of H${\alpha}$+[NII] line was estimated as $7500~\si{km.s^{-1}}$ from single Gaussian fitting. This FWHM is larger than the spectral resolution in the spectrum around this wavelength range $\simeq 3500~\si{km.s^{-1}}$ , indicating the presence of a broad H$\alpha$ component. However, since the contribution of [NII] is unclear, it remains an upper limit.  Although our estimated values are consistent with some faint AGNs, there are differences between these JWST-discovered faint AGNs \citep[see e.g.,][]{Maiolino23, Harikane23} and our target galaxies, HD1 and HD3. Potential H$\alpha$ broad line fluxes in our work are notably lower, approximately 1-dex lower than H$\alpha$ broad line flux of faint AGNs found by JWST in previous studies. Therefore, if HD1 and HD3 host AGNs, they must be very faint AGNs. This might explain the lack of other emission lines in HD1, but it is challenging to conclude definitively from our result. 

If we assume the line to be H$\alpha$+[NII] from star forming regions, we can derive the SFR by using the conversion factor of \citet{Kennicutt98} and a [NII]/H$\alpha$ ratio of $\sim1/3$ around the Solar metallicity \citep[e.g.,][]{2019A&ARv..27....3M}.
For HD1, $\text{SFR}_{\text{H}\alpha} \sim 0.4 M_\odot ~\text{yr}^{-1}$ after the dust correction based on $A_V$ from the SED fitting and the Calzetti law \citep{Calzetti00}. It is higher than the upper limit value of SFR derived from BAGPIPES; however, it is still consistent with a quiescent scenario. For HD3, the observed $\text{SFR}_{\text{H}\alpha}$ is approximately $0.5 M_{\odot} ~\text{yr}^{-1}$, which is higher than  $\text{SFR}_{\text{BAGPIPES}}=0.1 M_{\odot} ~\text{yr}^{-1}$. Although it exceeds the 2-dex lower bound of SFMS, it remains consistent with a quiescent scenario. These H$\alpha$ emission might result from present star formation. Additionally, it might indicate "rejuvenation" \citep[e.g.][]{Thomas10} where a galaxy experiences renewed star formation despite previous quiescent periods. Further investigation and analysis are essential to discern the true nature of this intriguing observation.

Secondly, we discuss about possibilities of a [NII] dominant case. Recently, a $z=2.445$ QG with an estimated stellar mass of $\log{(M_*/M_\odot)}={10.9}$ was discovered by \citet{Belli23}. They detected both [NII] and [OIII] emission lines in this QG. Based on the observed high [NII]/H${\alpha}$ and [OIII]/H${\beta}$ line ratios, they provided a compelling evidence for the presence of an AGN within the galaxy. Furthermore, their conclusions suggested that the AGN may have played a pivotal role in the quenching mechanism by expelling gas from the galaxy. In contrast, our observations of galaxy HD1 do not reveal H${\beta}$ or [OIII] emission lines, nor do they show any absorption lines of Ca II K and Na I D, which were observed in the target of \citet{Belli23}. The estimated H${\alpha}$+[NII] fluxes for HD1 and HD3 are notably weaker, approximately 2 dex lower than the [NII] flux observed in the galaxy studied by \citet{Belli23}. Note that the stellar masses of our targets are 0.5--1 dex lower than the QG in \citet{Belli23}. It is challenging to definitively determine whether this line is not an [NII] emission line or if the [OIII] lines are genuinely absent.

HD3 exhibits a potential [OIII] emission line feature, supporting the notion that HD3 has an AGN. However, the low resolution of our data prevents us from drawing definitive conclusions.

Certainly, there is a possibility that these H${\alpha}$+[NII] emission lines may result from a complex combination of these cases.

\subsection{Size-Mass relation}

The size-mass relation, describing the relationship between the effective radius and stellar mass of galaxies, is pivotal for comprehending their formation and evolution across cosmic time. Studying this relation across various galaxy populations, such as QGs and SFGs, enables us to discern differences in their evolutionary pathways.

\begin{figure*}
	\includegraphics[width=1\columnwidth, angle=270]{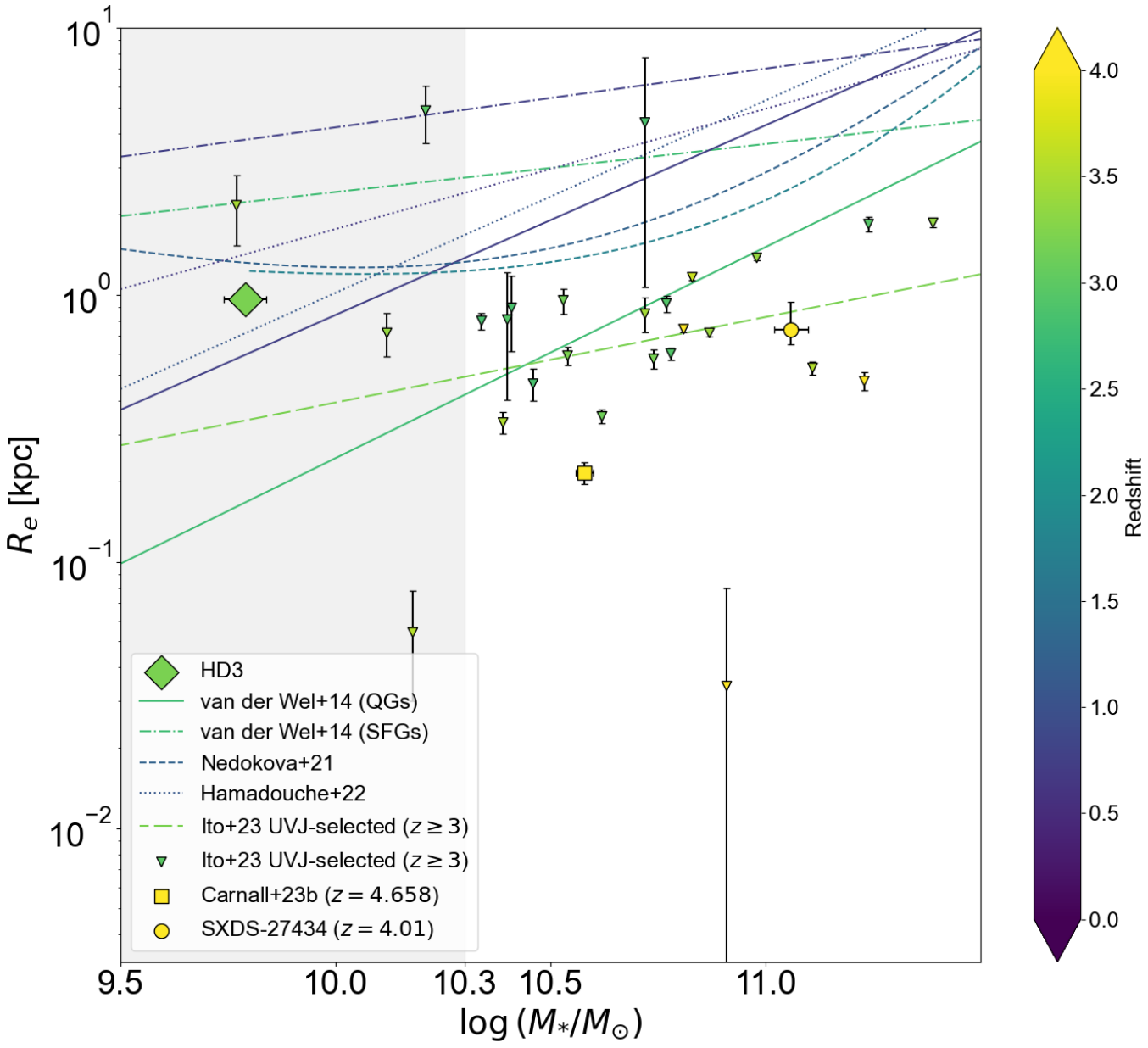}
    \caption{Size-mass relation and HD3. Each line shows size-mass relation of galaxies. \citet{Ito23} in long-dashed line ($UVJ$-selected). \citet{vanderWel14} in solid and dashed-dotted lines. \citet{Nedkova21} in dashed lines. \citet{Hamadouche22} in dotted lines. The data used to establish the size-mass relations in \citet{vanderWel14} only considered galaxies with masses exceeding ${10}^{10.3}M_\odot$, and the same criterion was followed in \citet{Hamadouche22, Ito23}. The region corresponds to $M_* < {10}^{10.3}M_\odot$ shown in grey-shaded region. Markers shows sizes and masses of individual quiescent galaxies from \citet{Carnall23b, Ito23} and SXDS-27434 \citep[see][]{Tanaka19, Valentino20, Ito23}. The colours of the symbols and lines indicate the redshift as shown in the colour bar. }
    \label{fig:size-mass}
\end{figure*}

Figure~\ref{fig:size-mass} shows the size-mass relation from some studies, including QGs and SFGs at $z \sim 0.75$ and $z \sim 2.75$ from \citet{vanderWel14}, QGs at  $z=0.6\text{--}0.8$ and $z=1.0\text{--}1.3$  from \citet{Hamadouche22}, and QGs at $z=1.0\text{--}1.5$ and $z=1.5\text{--}2.0$ from \citet{Nedkova21}, $UVJ$-selected QGs at $z\geq3$ from \citet{Ito23}. The relations from \citet{vanderWel14, Hamadouche22, Ito23} are extrapolated for stellar masses below ${10}^{10.3}M_\odot$. This is because the data used to establish the size-mass relations in \citet{vanderWel14} only considered galaxies with masses exceeding ${10}^{10.3}M_\odot$, and the same criterion was followed in \citet{Hamadouche22, Ito23}. It also depicts the sizes and masses of individual $UVJ$-selected QGs from \citet{Ito23}, as well as a QG at $z=4.658$ from \citet{Carnall23b}, SXDS-27434 ($z=4.01$) reported in \citet{Tanaka19} and \citet{Valentino20}, and HD3. 

In Figure~\ref{fig:size-mass}, HD3 occupies a position between what is typically observed for QGs and SFGs at $z=2.75$ as reported by \citet{vanderWel14}. Furthermore, HD3 falls above the extrapolated size-mass relation proposed by \citet{Ito23}. However, there is a larger scatter reported especially for the less massive regime of $\log{(M_*/M_\odot)} < 10.3$ \citep{Ito23,Cutler23} within which HD3 falls. 
Additionally, some studies \citep[e.g.][]{Kawinwanichakij21, Nedkova21} have reported bending in the size-mass relation of $z>2$ QGs at $\log{(M_*/M_\odot)} = 10 \text{--} 10.5$. This may explain HD3's position above the size-mass relation.

The S\'{e}rsic index of HD3 is $1.07 \pm 0.07$ as shown in Table~\ref{tab:hd3_morph}. This value is similar to some QGs in \citet{Ito23,Cutler23}, which display disc-like features. Suggesting that HD3 is currently undergoing a phase of morphological transition, akin to what \citet{Ito23,Cutler23} pointed out for their disc-like QG samples. \citet{Ito23} characterised a disc fraction of QGs, which is defined as the fraction of disc-like galaxies having the F277W S\'{e}rsic index below 2. They found a higher disc fraction of QGs in samples at $z > 3.3$ compared to those at $z < 3.3$. \citet{Cutler23} also indicate less-massive QG candidates are tends to have disc-like profiles based on photometric selected samples.

QGs show a steeper size-mass relation compared to SFGs in previous studies \citep[e.g.][]{vanderWel14}. High-redshift massive QGs are more compact and high dense within their effective radius comparing with their local counterparts. This alignment with inside-out growth scenarios, where minor mergers are implicated as origin, is supported by observations \citep[e.g.,][]{Bezanson09, vanDokkum10, Cimatti12, vanDokkum15} and simulations \citep[e.g.,][]{Naab09, Hilz13, Hopkins13}. \citet{Hamadouche22} reported that the size evolution of massive QGs from $z=1.1$ to $z=0.7$ might be dominated by minor mergers rather than major mergers. These minor mergers tend to make the galaxy radius larger, but the stellar mass does not evolve significantly when compared with major mergers, causing the size-mass relation to be steeper for QGs. However, HD3 is not showing elliptical in S\'{e}rsic profile and likely represents a disc population ($n=1$) with stellar mass lower than $\log{(M_*/M_\odot)} = 10$. \citet{Cutler23} reported a difference between the size-mass relation in less massive QG candidates and massive QG candidates, divided by $\log{(M_*/M_\odot)} = 10.3$. \citet{Cutler23} also found that massive QGs primarily evolve through mass quenching mechanisms, whereas less massive QGs are affected by other processes like environmental or feedback-driven quenching. Our findings support these variations in the mechanisms driving the evolution of galaxy size between massive QGs and less massive QGs. 

\subsection{Implications for quiescent galaxy quenching mechanism}

Considering quenching mechanisms, the timescale is an important factor to distinguish how quiescent galaxies (QGs) are quenched. There are roughly two types of quenching: "fast" and "slow" timescale quenching \citep[e.g.][]{Carnall18, Belli19}. From the result of SED fitting, we have quenching timescales of $\simeq 0.2\text{--} 0.7~\si{Gyr}$ for these galaxies, which are categorised as fast quenching. These values are consistent with quenching timescales for high-redshift QGs. \citet{Carnall18} concluded that rapid quenching with $< 1~\si{Gyr}$ might be the result of quasar-mode active galactic nucleus (AGN) feedback. In some simulations, AGN feedback is shown to contribute to rapid quenching. \citet{Weller24ax} used the cosmological simulations \texttt{IllustrisTNG} and \texttt{ASTRID}, measuring typical quenching timescales of approximately $\sim 200 \text{--} 600$ Myr for these two simulations. In both simulations, AGN feedback plays a crucial role in quenching star formation. However, in \texttt{IllustrisTNG} , AGN kinetic feedback is the dominant mechanism driving rapid quenching, whereas in \texttt{ASTRID}, AGN thermal feedback is less effective, leading to a slightly longer quenching timescale. Other simulations also conclude that AGN feedback plays a crucial role in quenching, but the mechanisms are complex and not solely dependent on AGN feedback \citep[e.g.,][]{Kimmig23ax, Remus23ax}. Combined with our discussion in Section~\ref{sec:origin} that HD1 and HD3 might host AGN, our target galaxies might be quenched by AGN feedback. 

It is also worth noting that we do not detect any companion galaxies in close proximity within a radius of $15~\si{arcsec}$ and redshift ranges $3.05$ to $3.35$ to HD3 in the JWST/PRIMER-UDS photo-z catalogue provided by DJA and CANDELS \citep{Grogin11, Koekemoer11}. HD1 and HD2 are positioned out of the area of CANDELS \citep{Grogin11, Koekemoer11}  COSMOS/UDS
photometric redshift catalogue \citep{Kodra23}, indicating the need for further research of environment around these objects. This observation raises questions about the quiescent galaxy formation scenario driven by environmental factors, as this scenario is often proposed for less massive quiescent galaxies. In the local universe, low-mass QGs are believed to be quenched by environmental effects \citep{Roberts19,Contini20}. \citet{Pasha23} noted that dwarf galaxies at $2<z<5$ with stellar masses ranging from $\log{(M_*/M_\odot)} = 5.5$ to $8.5$ are quenched by environmental effects such as shock heating from cosmic sheets and filaments, but given HD3's larger stellar mass, the influence of environmental effects remains uncertain.

It is important to note that the low dispersion of our current data and the absence of visible absorption lines prevent us from ruling out the possibility that the cessation of star formation may be temporarily due to strong supernova (SNe) feedback.

\subsection{Role of high redshift intermediate-mass QGs in galaxy formation and evolution}

Our result raises the possibility of the existence of AGNs in some quiescent galaxies (QGs) with a stellar mass range of $\log{(M_*/M_\odot)} < 11$. If this is the case, the progenitors of these galaxies might be explained by low-mass AGNs at higher redshift. In recent studies using JWST, unexpectedly large number of faint AGNs are found at $z \gtrsim 4$ (e.g. \citet{Harikane23, Maiolino23}). In \citet{Suh24}, a possible low-mass BH has found at $z \sim 4$. Studying intermediate-mass QGs might help our understanding of the descendants of these faint AGNs. By studying intermediate-mass QGs, we might be able to understand the relation between quenching and faint AGNs.

\citet{Weibel24ax} recently found a QG at $z \sim 7$ with similar stellar mass of our galaxies $\log{(M_*/M_\odot)} \sim 10.2$. Our galaxies, particularly HD1, may represent a similar type of QG, formed slightly later.  Additionally, a post-starburst galaxy at $z \sim 5$ \citep{Strait23} and a mini-quenched galaxy at $z \sim 7$ \citep{Looser24} have also been recently discovered by JWST. We need more samples to discuss the relations of QGs at higher redshifts $z>4$ with those at intermediate redshifts $z \lesssim 4$.

The intermediate-mass regime is another key feature of our sample, which is essential for determining which quenching mechanism is dominant depending on the stellar mass. Additionally, this regime helps clarify the role and position of QGs in the galaxy formation and evolution pathways, as well as their contribution to reionization and star formation across cosmic history.

\section{Conclusions}\label{sec:conclusion}

In this paper, we analyse JWST/NIRSpec spectrum of three galaxies HD1, HD2, and HD3 identified during the pre-JWST. Using \texttt{BAGPIPES} SED fitting code, together with available JWST/NIRCam photometry of HD3, we derive some physical parameters for these galaxies.

Our major findings in this fitting are as follows:
\begin{enumerate}
 \item For HD1, we find redshift $z=4.00$ from the position of the potentially-blended H${\alpha}$+[NII] line in its spectrum. HD1 is the least-massive QGs spectroscopically observed around $z=4$ with a stellar mass of $\log{(M_*/M_\odot)}={10.20}$.  The star formation rate (SFR) of HD1 is $\text{SFR}/M_\odot ~\text{yr}^{-1}< {0.12}$ which aligns with a quiescent scenario.
 \item HD2 is a less-massive QG at $z=3.29$ with a stellar mass of $\log{(M_*/M_\odot)}={10.16}$. The star formation rate (SFR) of HD2 is $\text{SFR}/M_\odot ~\text{yr}^{-1}= {0.00}$ which aligns with a quiescent scenario.
 \item  HD3 is the least-massive QG in our targets at $z=3.20$ with a stellar mass of $\log{(M_*/M_\odot)}={9.70}$. The star formation rate (SFR) of HD3 is $\text{SFR}/M_\odot ~\text{yr}^{-1}= {0.36}$ which aligns with a quiescent scenario.
\end{enumerate}

Two of the target galaxies, HD1 and HD3, exhibit a potential H${\alpha}$+[NII] emission line within their spectra. We discuss the possible origins of these emission lines. One possibility is that they are H$\alpha$-dominant emission lines from AGN \citep[e.g.][]{Carnall23b} within these galaxies or instantaneous star formation events like rejuvenation. Another possibility is that they are [NII]-dominant emission lines from AGN \citep[e.g.][]{Belli23} or a combination of these cases. Future high-resolution observations using JWST/NIRSpec or other instruments might provide more perspectives to this matter.

By using \texttt{Galfit} 2D profile fitting code, we study the morphology of HD3 in NIRCam images. HD3 exhibits a disc-like morphology with S\'{e}rsic index $n \simeq 1$. From this result, HD3 may be undergoing a phase of morphological transformation. HD3 has larger effective radius expected by extrapolated size-mass relation of $z\sim3$ QGs in \citet{vanderWel14, Ito23}  This result is consistent with flatten of size-mass relation of QGs in less-massive end reported in \citet{Kawinwanichakij21, Nedkova21}. This might be result from difference in size evolution pathways between high-mass QGs and low-mass QGs \citep[e.g.][]{vanDokkum15,Cutler23}.

Recent observations with JWST \citep{Nanayakkara24, Carnall23b} and simulations \citep[][]{Hartley23,Bluck24} suggest that AGNs are the most likely progenitors for the quenching of massive QGs ($\log{(M_/M_\odot)} > 10$) at $z>3$. However, for low-mass QGs ($\log{(M_/M_\odot)} < 9$), environmental quenching is considered a possible mechanism. Nonetheless, the quenching mechanisms of high-redshift QGs with stellar masses around $\log{(M_/M_\odot)} \simeq 9\texttt{--}10$ remain uncertain. The short quenching timescales of our targets, ranging from $0.2$ to $0.7~\si{Gyr}$, suggest that these galaxies might be quenched by strong AGN feedback \citep{Carnall18}, combined with the possible presence of AGN indicated from H${\alpha}$+[NII] emission line in HD1 and HD3.
The lack of detected companion galaxies in close proximity to HD3 challenges the prevailing idea that environmental factors drive the formation of quiescent galaxies, particularly for less massive ones with stellar masses of $\sim \log{(M_*/M_\odot)}={10}$.  If AGNs exist in our targets, the possibility of AGN-driven quenching in the intermediate mass range of $\log{(M_*/M_\odot)} \simeq 9\texttt{--}10$ becomes more plausible. Future deep and high-resolution observations of our targets and other intermediate-mass QGs by NIRSpec and NIRCam or other instruments could be crucial in exploring the quenching mechanisms and understanding the galaxy formation and evolution pathway across the cosmic history.

\section*{Acknowledgements}

We are grateful for useful discussion with Takumi Tanaka. We are also thankful for the valuable information regarding the NIRSpec data and for the support and expertise in NIRSpec and data reduction offered by the JWST helpdesk, especially Charles R. Proffitt, Tony Keyes, and Maria Pena-Guerrero.

We deeply thank the anonymous referee for constructive comments that improve the quality and clarity of this paper.

AKI is supported by JSPS KAKENHI Grant Number 23H00131.
YT is supported by JSPS KAKENHI Grant Number 22H04939 and 23K20035.
T.H. was supported by Leading Initiative for Excellent Young Researchers, MEXT, Japan (HJH02007) and by JSPS KAKENHI grant No. 22H01258.
K.I. was supported by JSPS KAKENHI Grant Numbers JP22J00495, and JP23K13141.

This work is based on observations made with the NASA/ESA/CSA James Webb Space Telescope. The data were obtained from the Mikulski Archive for Space Telescopes at the Space Telescope Science Institute, which is operated by the Association of Universities for Research in Astronomy, Inc., under NASA contract NAS 5-03127 for JWST. These observations are associated with program $\#1740$.
This paper uses the data products retrieved from the Dawn JWST Archive (DJA). DJA is an initiative of the Cosmic Dawn Center, which is funded by the Danish National Research Foundation under grant No. 140.

\section*{Data Availability}

All NIRSpec data are available via the \texttt{Mikulski Archive for Space Telescopes (MAST)} (\url{https://mast.stsci.edu}).
The \texttt{Bagpipes} code is publicly available at \url{https://github.com/ACCarnall/bagpipes}.
NIRCam and MIRI imaging data used in this paper are available in the Dawn JWST Archive (DJA) at \url{https://dawn-cph.github.io/dja/}. 



\bibliographystyle{mnras}
\bibliography{main} 

\begin{thebibliography}{}
\makeatletter
\relax
\def\mn@urlcharsother{\let\do\@makeother \do\$\do\&\do\#\do\^\do\_\do\%\do\~}
\def\mn@doi{\begingroup\mn@urlcharsother \@ifnextchar [ {\mn@doi@} {\mn@doi@[]}}
\def\mn@doi@[#1]#2{\def\@tempa{#1}\ifx\@tempa\@empty \href {http://dx.doi.org/#2} {doi:#2}\else \href {http://dx.doi.org/#2} {#1}\fi \endgroup}
\def\mn@eprint#1#2{\mn@eprint@#1:#2::\@nil}
\def\mn@eprint@arXiv#1{\href {http://arxiv.org/abs/#1} {{\tt arXiv:#1}}}
\def\mn@eprint@dblp#1{\href {http://dblp.uni-trier.de/rec/bibtex/#1.xml} {dblp:#1}}
\def\mn@eprint@#1:#2:#3:#4\@nil{\def\@tempa {#1}\def\@tempb {#2}\def\@tempc {#3}\ifx \@tempc \@empty \let \@tempc \@tempb \let \@tempb \@tempa \fi \ifx \@tempb \@empty \def\@tempb {arXiv}\fi \@ifundefined {mn@eprint@\@tempb}{\@tempb:\@tempc}{\expandafter \expandafter \csname mn@eprint@\@tempb\endcsname \expandafter{\@tempc}}}

\bibitem[\protect\citeauthoryear{{Aihara} et~al.,}{{Aihara} et~al.}{2018}]{Aihara18}
{Aihara} H.,  et~al., 2018, \mn@doi [\pasj] {10.1093/pasj/psx066}, \href {https://ui.adsabs.harvard.edu/abs/2018PASJ...70S...4A} {70, S4}

\bibitem[\protect\citeauthoryear{{Aihara} et~al.,}{{Aihara} et~al.}{2019}]{Aihara19}
{Aihara} H.,  et~al., 2019, \mn@doi [\pasj] {10.1093/pasj/psz103}, \href {https://ui.adsabs.harvard.edu/abs/2019PASJ...71..114A} {71, 114}

\bibitem[\protect\citeauthoryear{{Alberts} et~al.,}{{Alberts} et~al.}{2023}]{Alberts23}
{Alberts} S.,  et~al., 2023, \mn@doi [arXiv e-prints] {10.48550/arXiv.2312.12207}, \href {https://ui.adsabs.harvard.edu/abs/2023arXiv231212207A} {p. arXiv:2312.12207}

\bibitem[\protect\citeauthoryear{{Antwi-Danso} et~al.,}{{Antwi-Danso} et~al.}{2023}]{Antwi-Danso23}
{Antwi-Danso} J.,  et~al., 2023, \mn@doi [arXiv e-prints] {10.48550/arXiv.2307.09590}, \href {https://ui.adsabs.harvard.edu/abs/2023arXiv230709590A} {p. arXiv:2307.09590}

\bibitem[\protect\citeauthoryear{{Belli}, {Newman}  \& {Ellis}}{{Belli} et~al.}{2019}]{Belli19}
{Belli} S.,  {Newman} A.~B.,   {Ellis} R.~S.,  2019, \mn@doi [\apj] {10.3847/1538-4357/ab07af}, \href {https://ui.adsabs.harvard.edu/abs/2019ApJ...874...17B} {874, 17}

\bibitem[\protect\citeauthoryear{{Belli} et~al.,}{{Belli} et~al.}{2023}]{Belli23}
{Belli} S.,  et~al., 2023, \mn@doi [arXiv e-prints] {10.48550/arXiv.2308.05795}, \href {https://ui.adsabs.harvard.edu/abs/2023arXiv230805795B} {p. arXiv:2308.05795}

\bibitem[\protect\citeauthoryear{{Bezanson}, {van Dokkum}, {Tal}, {Marchesini}, {Kriek}, {Franx}  \& {Coppi}}{{Bezanson} et~al.}{2009}]{Bezanson09}
{Bezanson} R.,  {van Dokkum} P.~G.,  {Tal} T.,  {Marchesini} D.,  {Kriek} M.,  {Franx} M.,   {Coppi} P.,  2009, \mn@doi [\apj] {10.1088/0004-637X/697/2/1290}, \href {https://ui.adsabs.harvard.edu/abs/2009ApJ...697.1290B} {697, 1290}

\bibitem[\protect\citeauthoryear{{Bluck} et~al.,}{{Bluck} et~al.}{2023}]{Bluck23b}
{Bluck} A. F.~L.,  et~al., 2023, \mn@doi [arXiv e-prints] {10.48550/arXiv.2311.02526}, \href {https://ui.adsabs.harvard.edu/abs/2023arXiv231102526B} {p. arXiv:2311.02526}

\bibitem[\protect\citeauthoryear{{Bluck} et~al.,}{{Bluck} et~al.}{2024}]{Bluck24}
{Bluck} A. F.~L.,  et~al., 2024, \mn@doi [\apj] {10.3847/1538-4357/ad0a98}, \href {https://ui.adsabs.harvard.edu/abs/2024ApJ...961..163B} {961, 163}

\bibitem[\protect\citeauthoryear{{Boselli}, {Fossati}  \& {Sun}}{{Boselli} et~al.}{2022}]{Boselli22}
{Boselli} A.,  {Fossati} M.,   {Sun} M.,  2022, \mn@doi [\aapr] {10.1007/s00159-022-00140-3}, \href {https://ui.adsabs.harvard.edu/abs/2022A&ARv..30....3B} {30, 3}

\bibitem[\protect\citeauthoryear{{Bradley} et~al.,}{{Bradley} et~al.}{2022}]{photutils}
{Bradley} L.,  et~al., 2022, {astropy/photutils: 1.5.0}, Zenodo, \mn@doi{10.5281/zenodo.6825092}

\bibitem[\protect\citeauthoryear{{Brinchmann}, {Charlot}, {White}, {Tremonti}, {Kauffmann}, {Heckman}  \& {Brinkmann}}{{Brinchmann} et~al.}{2004}]{Brinchmann04}
{Brinchmann} J.,  {Charlot} S.,  {White} S.~D.~M.,  {Tremonti} C.,  {Kauffmann} G.,  {Heckman} T.,   {Brinkmann} J.,  2004, \mn@doi [\mnras] {10.1111/j.1365-2966.2004.07881.x}, \href {https://ui.adsabs.harvard.edu/abs/2004MNRAS.351.1151B} {351, 1151}

\bibitem[\protect\citeauthoryear{{Bruzual} \& {Charlot}}{{Bruzual} \& {Charlot}}{2003}]{BC03}
{Bruzual} G.,  {Charlot} S.,  2003, \mn@doi [\mnras] {10.1046/j.1365-8711.2003.06897.x}, \href {https://ui.adsabs.harvard.edu/abs/2003MNRAS.344.1000B} {344, 1000}

\bibitem[\protect\citeauthoryear{{Calzetti}, {Armus}, {Bohlin}, {Kinney}, {Koornneef}  \& {Storchi-Bergmann}}{{Calzetti} et~al.}{2000}]{Calzetti00}
{Calzetti} D.,  {Armus} L.,  {Bohlin} R.~C.,  {Kinney} A.~L.,  {Koornneef} J.,   {Storchi-Bergmann} T.,  2000, \mn@doi [\apj] {10.1086/308692}, \href {https://ui.adsabs.harvard.edu/abs/2000ApJ...533..682C} {533, 682}

\bibitem[\protect\citeauthoryear{{Carnall}, {McLure}, {Dunlop}  \& {Dav{\'e}}}{{Carnall} et~al.}{2018}]{Carnall18}
{Carnall} A.~C.,  {McLure} R.~J.,  {Dunlop} J.~S.,   {Dav{\'e}} R.,  2018, \mn@doi [\mnras] {10.1093/mnras/sty2169}, \href {https://ui.adsabs.harvard.edu/abs/2018MNRAS.480.4379C} {480, 4379}

\bibitem[\protect\citeauthoryear{{Carnall} et~al.,}{{Carnall} et~al.}{2023a}]{Carnall23a}
{Carnall} A.~C.,  et~al., 2023a, \mn@doi [\mnras] {10.1093/mnras/stad369}, \href {https://ui.adsabs.harvard.edu/abs/2023MNRAS.520.3974C} {520, 3974}

\bibitem[\protect\citeauthoryear{Carnall et~al.,}{Carnall et~al.}{2023b}]{Carnall23b}
Carnall A.~C.,  et~al., 2023b, \mn@doi [Nature] {10.1038/s41586-023-06158-6}, \href {https://ui.adsabs.harvard.edu/abs/2023Natur.619..716C} {619, 716–719}

\bibitem[\protect\citeauthoryear{{Carnall} et~al.,}{{Carnall} et~al.}{2024}]{Carnall24}
{Carnall} A.~C.,  et~al., 2024, \mn@doi [arXiv e-prints] {10.48550/arXiv.2405.02242}, \href {https://ui.adsabs.harvard.edu/abs/2024arXiv240502242C} {p. arXiv:2405.02242}

\bibitem[\protect\citeauthoryear{{Chevallard} \& {Charlot}}{{Chevallard} \& {Charlot}}{2016}]{BC16}
{Chevallard} J.,  {Charlot} S.,  2016, \mn@doi [\mnras] {10.1093/mnras/stw1756}, \href {https://ui.adsabs.harvard.edu/abs/2016MNRAS.462.1415C} {462, 1415}

\bibitem[\protect\citeauthoryear{{Cimatti}, {Nipoti}  \& {Cassata}}{{Cimatti} et~al.}{2012}]{Cimatti12}
{Cimatti} A.,  {Nipoti} C.,   {Cassata} P.,  2012, \mn@doi [\mnras] {10.1111/j.1745-3933.2012.01237.x}, \href {https://ui.adsabs.harvard.edu/abs/2012MNRAS.422L..62C} {422, L62}

\bibitem[\protect\citeauthoryear{{Contini}, {Gu}, {Ge}, {Rhee}, {Yi}  \& {Kang}}{{Contini} et~al.}{2020}]{Contini20}
{Contini} E.,  {Gu} Q.,  {Ge} X.,  {Rhee} J.,  {Yi} S.~K.,   {Kang} X.,  2020, \mn@doi [\apj] {10.3847/1538-4357/ab6730}, \href {https://ui.adsabs.harvard.edu/abs/2020ApJ...889..156C} {889, 156}

\bibitem[\protect\citeauthoryear{{Croton} et~al.,}{{Croton} et~al.}{2006}]{Croton06}
{Croton} D.~J.,  et~al., 2006, \mn@doi [\mnras] {10.1111/j.1365-2966.2005.09675.x}, \href {https://ui.adsabs.harvard.edu/abs/2006MNRAS.365...11C} {365, 11}

\bibitem[\protect\citeauthoryear{{Cutler} et~al.,}{{Cutler} et~al.}{2023}]{Cutler23}
{Cutler} S.~E.,  et~al., 2023, \mn@doi [arXiv e-prints] {10.48550/arXiv.2312.15012}, \href {https://ui.adsabs.harvard.edu/abs/2023arXiv231215012C} {p. arXiv:2312.15012}

\bibitem[\protect\citeauthoryear{{D'Eugenio} et~al.,}{{D'Eugenio} et~al.}{2021}]{D'Eugenio21}
{D'Eugenio} C.,  et~al., 2021, \mn@doi [\aap] {10.1051/0004-6361/202040067}, \href {https://ui.adsabs.harvard.edu/abs/2021A&A...653A..32D} {653, A32}

\bibitem[\protect\citeauthoryear{{Davidzon} et~al.,}{{Davidzon} et~al.}{2017}]{Davidzon17}
{Davidzon} I.,  et~al., 2017, \mn@doi [\aap] {10.1051/0004-6361/201730419}, \href {https://ui.adsabs.harvard.edu/abs/2017A&A...605A..70D} {605, A70}

\bibitem[\protect\citeauthoryear{Earl et~al.,}{Earl et~al.}{2023}]{spectutils}
Earl N.,  et~al., 2023, astropy/specutils: v1.11.0, \mn@doi{10.5281/zenodo.8049033}, \url {https://doi.org/10.5281/zenodo.8049033}

\bibitem[\protect\citeauthoryear{{Forrest} et~al.,}{{Forrest} et~al.}{2020}]{Forrest20}
{Forrest} B.,  et~al., 2020, \mn@doi [\apjl] {10.3847/2041-8213/ab5b9f}, \href {https://ui.adsabs.harvard.edu/abs/2020ApJ...890L...1F} {890, L1}

\bibitem[\protect\citeauthoryear{{Furusawa} et~al.,}{{Furusawa} et~al.}{2008}]{Furusawa08}
{Furusawa} H.,  et~al., 2008, \mn@doi [\apjs] {10.1086/527321}, \href {https://ui.adsabs.harvard.edu/abs/2008ApJS..176....1F} {176, 1}

\bibitem[\protect\citeauthoryear{{Gabor}, {Dav{\'e}}, {Finlator}  \& {Oppenheimer}}{{Gabor} et~al.}{2010}]{Gabor10}
{Gabor} J.~M.,  {Dav{\'e}} R.,  {Finlator} K.,   {Oppenheimer} B.~D.,  2010, \mn@doi [\mnras] {10.1111/j.1365-2966.2010.16961.x}, \href {https://ui.adsabs.harvard.edu/abs/2010MNRAS.407..749G} {407, 749}

\bibitem[\protect\citeauthoryear{{Glazebrook} et~al.,}{{Glazebrook} et~al.}{2017}]{Glazebrook17}
{Glazebrook} K.,  et~al., 2017, \mn@doi [\nat] {10.1038/nature21680}, \href {https://ui.adsabs.harvard.edu/abs/2017Natur.544...71G} {544, 71}

\bibitem[\protect\citeauthoryear{{Glazebrook} et~al.,}{{Glazebrook} et~al.}{2024}]{Glazebrook24}
{Glazebrook} K.,  et~al., 2024, \mn@doi [\nat] {10.1038/s41586-024-07191-9}, \href {https://ui.adsabs.harvard.edu/abs/2024Natur.628..277G} {628, 277}

\bibitem[\protect\citeauthoryear{{Greene} \& {Ho}}{{Greene} \& {Ho}}{2005}]{GreeneHo05}
{Greene} J.~E.,  {Ho} L.~C.,  2005, \mn@doi [\apj] {10.1086/431897}, \href {https://ui.adsabs.harvard.edu/abs/2005ApJ...630..122G} {630, 122}

\bibitem[\protect\citeauthoryear{{Grogin} et~al.,}{{Grogin} et~al.}{2011}]{Grogin11}
{Grogin} N.~A.,  et~al., 2011, \mn@doi [\apjs] {10.1088/0067-0049/197/2/35}, \href {https://ui.adsabs.harvard.edu/abs/2011ApJS..197...35G} {197, 35}

\bibitem[\protect\citeauthoryear{{Hamadouche} et~al.,}{{Hamadouche} et~al.}{2022}]{Hamadouche22}
{Hamadouche} M.~L.,  et~al., 2022, \mn@doi [\mnras] {10.1093/mnras/stac535}, \href {https://ui.adsabs.harvard.edu/abs/2022MNRAS.512.1262H} {512, 1262}

\bibitem[\protect\citeauthoryear{{Harikane} et~al.,}{{Harikane} et~al.}{2022}]{Harikane22a}
{Harikane} Y.,  et~al., 2022, \mn@doi [\apj] {10.3847/1538-4357/ac53a9}, \href {https://ui.adsabs.harvard.edu/abs/2022ApJ...929....1H} {929, 1}

\bibitem[\protect\citeauthoryear{{Harikane} et~al.,}{{Harikane} et~al.}{2023}]{Harikane23}
{Harikane} Y.,  et~al., 2023, \mn@doi [\apj] {10.3847/1538-4357/ad029e}, \href {https://ui.adsabs.harvard.edu/abs/2023ApJ...959...39H} {959, 39}

\bibitem[\protect\citeauthoryear{{Harikane} et~al.,}{{Harikane} et~al.}{2024}]{Harikane24}
{Harikane} Y.,  et~al., 2024, \mn@doi [arXiv e-prints] {10.48550/arXiv.2406.18352}, \href {https://ui.adsabs.harvard.edu/abs/2024arXiv240618352H} {p. arXiv:2406.18352}

\bibitem[\protect\citeauthoryear{{Hartley} et~al.,}{{Hartley} et~al.}{2023}]{Hartley23}
{Hartley} A.~I.,  et~al., 2023, \mn@doi [\mnras] {10.1093/mnras/stad1162}, \href {https://ui.adsabs.harvard.edu/abs/2023MNRAS.522.3138H} {522, 3138}

\bibitem[\protect\citeauthoryear{{Hilz}, {Naab}  \& {Ostriker}}{{Hilz} et~al.}{2013}]{Hilz13}
{Hilz} M.,  {Naab} T.,   {Ostriker} J.~P.,  2013, \mn@doi [\mnras] {10.1093/mnras/sts501}, \href {https://ui.adsabs.harvard.edu/abs/2013MNRAS.429.2924H} {429, 2924}

\bibitem[\protect\citeauthoryear{{Hopkins}, {Cox}, {Hernquist}, {Narayanan}, {Hayward}  \& {Murray}}{{Hopkins} et~al.}{2013}]{Hopkins13}
{Hopkins} P.~F.,  {Cox} T.~J.,  {Hernquist} L.,  {Narayanan} D.,  {Hayward} C.~C.,   {Murray} N.,  2013, \mn@doi [\mnras] {10.1093/mnras/stt017}, \href {https://ui.adsabs.harvard.edu/abs/2013MNRAS.430.1901H} {430, 1901}

\bibitem[\protect\citeauthoryear{{Houston}, {Croton}  \& {Sinha}}{{Houston} et~al.}{2023}]{Houston23}
{Houston} T.,  {Croton} D.~J.,   {Sinha} M.,  2023, \mn@doi [\mnras] {10.1093/mnrasl/slad031}, \href {https://ui.adsabs.harvard.edu/abs/2023MNRAS.522L..11H} {522, L11}

\bibitem[\protect\citeauthoryear{{Ito} et~al.,}{{Ito} et~al.}{2023}]{Ito23}
{Ito} K.,  et~al., 2023, \mn@doi [arXiv e-prints] {10.48550/arXiv.2307.06994}, \href {https://ui.adsabs.harvard.edu/abs/2023arXiv230706994I} {p. arXiv:2307.06994}

\bibitem[\protect\citeauthoryear{{Kakimoto} et~al.,}{{Kakimoto} et~al.}{2024}]{Kakimoto24}
{Kakimoto} T.,  et~al., 2024, \mn@doi [\apj] {10.3847/1538-4357/ad1ff1}, \href {https://ui.adsabs.harvard.edu/abs/2024ApJ...963...49K} {963, 49}

\bibitem[\protect\citeauthoryear{{Kawinwanichakij} et~al.,}{{Kawinwanichakij} et~al.}{2021}]{Kawinwanichakij21}
{Kawinwanichakij} L.,  et~al., 2021, \mn@doi [\apj] {10.3847/1538-4357/ac1f21}, \href {https://ui.adsabs.harvard.edu/abs/2021ApJ...921...38K} {921, 38}

\bibitem[\protect\citeauthoryear{{Kennicutt}}{{Kennicutt}}{1998}]{Kennicutt98}
{Kennicutt} Robert~C. J.,  1998, \mn@doi [\araa] {10.1146/annurev.astro.36.1.189}, \href {https://ui.adsabs.harvard.edu/abs/1998ARA&A..36..189K} {36, 189}

\bibitem[\protect\citeauthoryear{{Kimmig}, {Remus}, {Seidel}, {Valenzuela}, {Dolag}  \& {Burkert}}{{Kimmig} et~al.}{2023}]{Kimmig23ax}
{Kimmig} L.~C.,  {Remus} R.-S.,  {Seidel} B.,  {Valenzuela} L.~M.,  {Dolag} K.,   {Burkert} A.,  2023, \mn@doi [arXiv e-prints] {10.48550/arXiv.2310.16085}, \href {https://ui.adsabs.harvard.edu/abs/2023arXiv231016085K} {p. arXiv:2310.16085}

\bibitem[\protect\citeauthoryear{{Kodra} et~al.,}{{Kodra} et~al.}{2023}]{Kodra23}
{Kodra} D.,  et~al., 2023, \mn@doi [\apj] {10.3847/1538-4357/ac9f12}, \href {https://ui.adsabs.harvard.edu/abs/2023ApJ...942...36K} {942, 36}

\bibitem[\protect\citeauthoryear{{Koekemoer} et~al.,}{{Koekemoer} et~al.}{2011}]{Koekemoer11}
{Koekemoer} A.~M.,  et~al., 2011, \mn@doi [\apjs] {10.1088/0067-0049/197/2/36}, \href {https://ui.adsabs.harvard.edu/abs/2011ApJS..197...36K} {197, 36}

\bibitem[\protect\citeauthoryear{{Kroupa} \& {Boily}}{{Kroupa} \& {Boily}}{2002}]{Kroupa02}
{Kroupa} P.,  {Boily} C.~M.,  2002, \mn@doi [\mnras] {10.1046/j.1365-8711.2002.05848.x}, \href {https://ui.adsabs.harvard.edu/abs/2002MNRAS.336.1188K} {336, 1188}

\bibitem[\protect\citeauthoryear{{Lawrence} et~al.,}{{Lawrence} et~al.}{2007}]{Lawrence07}
{Lawrence} A.,  et~al., 2007, \mn@doi [\mnras] {10.1111/j.1365-2966.2007.12040.x}, \href {https://ui.adsabs.harvard.edu/abs/2007MNRAS.379.1599L} {379, 1599}

\bibitem[\protect\citeauthoryear{{Looser} et~al.,}{{Looser} et~al.}{2024}]{Looser24}
{Looser} T.~J.,  et~al., 2024, \mn@doi [\nat] {10.1038/s41586-024-07227-0}, \href {https://ui.adsabs.harvard.edu/abs/2024Natur.629...53L} {629, 53}

\bibitem[\protect\citeauthoryear{{Maiolino} \& {Mannucci}}{{Maiolino} \& {Mannucci}}{2019}]{2019A&ARv..27....3M}
{Maiolino} R.,  {Mannucci} F.,  2019, \mn@doi [\aapr] {10.1007/s00159-018-0112-2}, \href {https://ui.adsabs.harvard.edu/abs/2019A&ARv..27....3M} {27, 3}

\bibitem[\protect\citeauthoryear{{Maiolino} et~al.,}{{Maiolino} et~al.}{2012}]{Maiolino12}
{Maiolino} R.,  et~al., 2012, \mn@doi [\mnras] {10.1111/j.1745-3933.2012.01303.x}, \href {https://ui.adsabs.harvard.edu/abs/2012MNRAS.425L..66M} {425, L66}

\bibitem[\protect\citeauthoryear{{Maiolino} et~al.,}{{Maiolino} et~al.}{2023}]{Maiolino23}
{Maiolino} R.,  et~al., 2023, \mn@doi [arXiv e-prints] {10.48550/arXiv.2308.01230}, \href {https://ui.adsabs.harvard.edu/abs/2023arXiv230801230M} {p. arXiv:2308.01230}

\bibitem[\protect\citeauthoryear{{Marchesini} et~al.,}{{Marchesini} et~al.}{2023}]{Marchesini23}
{Marchesini} D.,  et~al., 2023, \mn@doi [\apjl] {10.3847/2041-8213/acaaac}, \href {https://ui.adsabs.harvard.edu/abs/2023ApJ...942L..25M} {942, L25}

\bibitem[\protect\citeauthoryear{{Martig}, {Bournaud}, {Teyssier}  \& {Dekel}}{{Martig} et~al.}{2009}]{Martig09}
{Martig} M.,  {Bournaud} F.,  {Teyssier} R.,   {Dekel} A.,  2009, \mn@doi [\apj] {10.1088/0004-637X/707/1/250}, \href {https://ui.adsabs.harvard.edu/abs/2009ApJ...707..250M} {707, 250}

\bibitem[\protect\citeauthoryear{{Mawatari}, {Yamada}, {Fazio}, {Huang}  \& {Ashby}}{{Mawatari} et~al.}{2016}]{Mawatari16}
{Mawatari} K.,  {Yamada} T.,  {Fazio} G.~G.,  {Huang} J.-S.,   {Ashby} M. L.~N.,  2016, \mn@doi [\pasj] {10.1093/pasj/psw041}, \href {https://ui.adsabs.harvard.edu/abs/2016PASJ...68...46M} {68, 46}

\bibitem[\protect\citeauthoryear{{Mawatari} et~al.,}{{Mawatari} et~al.}{2020}]{Mawatari20}
{Mawatari} K.,  et~al., 2020, \mn@doi [\apj] {10.3847/1538-4357/ab6596}, \href {https://ui.adsabs.harvard.edu/abs/2020ApJ...889..137M} {889, 137}

\bibitem[\protect\citeauthoryear{{McCracken} et~al.,}{{McCracken} et~al.}{2012}]{McCracken12}
{McCracken} H.~J.,  et~al., 2012, \mn@doi [\aap] {10.1051/0004-6361/201219507}, \href {https://ui.adsabs.harvard.edu/abs/2012A&A...544A.156M} {544, A156}

\bibitem[\protect\citeauthoryear{{Merlin} et~al.,}{{Merlin} et~al.}{2019}]{Merlin19}
{Merlin} E.,  et~al., 2019, \mn@doi [\mnras] {10.1093/mnras/stz2615}, \href {https://ui.adsabs.harvard.edu/abs/2019MNRAS.490.3309M} {490, 3309}

\bibitem[\protect\citeauthoryear{{Naab}, {Johansson}  \& {Ostriker}}{{Naab} et~al.}{2009}]{Naab09}
{Naab} T.,  {Johansson} P.~H.,   {Ostriker} J.~P.,  2009, \mn@doi [\apjl] {10.1088/0004-637X/699/2/L178}, \href {https://ui.adsabs.harvard.edu/abs/2009ApJ...699L.178N} {699, L178}

\bibitem[\protect\citeauthoryear{{Nanayakkara} et~al.,}{{Nanayakkara} et~al.}{2024}]{Nanayakkara24}
{Nanayakkara} T.,  et~al., 2024, \mn@doi [Scientific Reports] {10.1038/s41598-024-52585-4}, \href {https://ui.adsabs.harvard.edu/abs/2024NatSR..14.3724N} {14, 3724}

\bibitem[\protect\citeauthoryear{{Nedkova} et~al.,}{{Nedkova} et~al.}{2021}]{Nedkova21}
{Nedkova} K.~V.,  et~al., 2021, \mn@doi [\mnras] {10.1093/mnras/stab1744}, \href {https://ui.adsabs.harvard.edu/abs/2021MNRAS.506..928N} {506, 928}

\bibitem[\protect\citeauthoryear{{Nelson} et~al.,}{{Nelson} et~al.}{2012}]{Nelson12}
{Nelson} E.~J.,  et~al., 2012, \mn@doi [\apjl] {10.1088/2041-8205/747/2/L28}, \href {https://ui.adsabs.harvard.edu/abs/2012ApJ...747L..28N} {747, L28}

\bibitem[\protect\citeauthoryear{{Nelson} et~al.,}{{Nelson} et~al.}{2016}]{Nelson16}
{Nelson} E.~J.,  et~al., 2016, \mn@doi [\apj] {10.3847/0004-637X/828/1/27}, \href {https://ui.adsabs.harvard.edu/abs/2016ApJ...828...27N} {828, 27}

\bibitem[\protect\citeauthoryear{{Newman}, {Ellis}, {Bundy}  \& {Treu}}{{Newman} et~al.}{2012}]{Newman12}
{Newman} A.~B.,  {Ellis} R.~S.,  {Bundy} K.,   {Treu} T.,  2012, \mn@doi [\apj] {10.1088/0004-637X/746/2/162}, \href {https://ui.adsabs.harvard.edu/abs/2012ApJ...746..162N} {746, 162}

\bibitem[\protect\citeauthoryear{{Ownsworth}, {Conselice}, {Mortlock}, {Hartley}, {Almaini}, {Duncan}  \& {Mundy}}{{Ownsworth} et~al.}{2014}]{Ownsworth14}
{Ownsworth} J.~R.,  {Conselice} C.~J.,  {Mortlock} A.,  {Hartley} W.~G.,  {Almaini} O.,  {Duncan} K.,   {Mundy} C.~J.,  2014, \mn@doi [\mnras] {10.1093/mnras/stu1802}, \href {https://ui.adsabs.harvard.edu/abs/2014MNRAS.445.2198O} {445, 2198}

\bibitem[\protect\citeauthoryear{{Pasha}, {Mandelker}, {van den Bosch}, {Springel}  \& {van de Voort}}{{Pasha} et~al.}{2023}]{Pasha23}
{Pasha} I.,  {Mandelker} N.,  {van den Bosch} F.~C.,  {Springel} V.,   {van de Voort} F.,  2023, \mn@doi [\mnras] {10.1093/mnras/stac3776}, \href {https://ui.adsabs.harvard.edu/abs/2023MNRAS.520.2692P} {520, 2692}

\bibitem[\protect\citeauthoryear{{Peng}, {Ho}, {Impey}  \& {Rix}}{{Peng} et~al.}{2002}]{Peng02}
{Peng} C.~Y.,  {Ho} L.~C.,  {Impey} C.~D.,   {Rix} H.-W.,  2002, \mn@doi [\aj] {10.1086/340952}, \href {https://ui.adsabs.harvard.edu/abs/2002AJ....124..266P} {124, 266}

\bibitem[\protect\citeauthoryear{{Peng}, {Ho}, {Impey}  \& {Rix}}{{Peng} et~al.}{2010a}]{Peng10}
{Peng} C.~Y.,  {Ho} L.~C.,  {Impey} C.~D.,   {Rix} H.-W.,  2010a, \mn@doi [\aj] {10.1088/0004-6256/139/6/2097}, \href {https://ui.adsabs.harvard.edu/abs/2010AJ....139.2097P} {139, 2097}

\bibitem[\protect\citeauthoryear{{Peng} et~al.,}{{Peng} et~al.}{2010b}]{YJPeng10}
{Peng} Y.-j.,  et~al., 2010b, \mn@doi [\apj] {10.1088/0004-637X/721/1/193}, \href {https://ui.adsabs.harvard.edu/abs/2010ApJ...721..193P} {721, 193}

\bibitem[\protect\citeauthoryear{{Peng}, {Maiolino}  \& {Cochrane}}{{Peng} et~al.}{2015}]{Peng15}
{Peng} Y.,  {Maiolino} R.,   {Cochrane} R.,  2015, \mn@doi [\nat] {10.1038/nature14439}, \href {https://ui.adsabs.harvard.edu/abs/2015Natur.521..192P} {521, 192}

\bibitem[\protect\citeauthoryear{{Piotrowska}, {Bluck}, {Maiolino}  \& {Peng}}{{Piotrowska} et~al.}{2022}]{Piotrowska22}
{Piotrowska} J.~M.,  {Bluck} A. F.~L.,  {Maiolino} R.,   {Peng} Y.,  2022, \mn@doi [\mnras] {10.1093/mnras/stab3673}, \href {https://ui.adsabs.harvard.edu/abs/2022MNRAS.512.1052P} {512, 1052}

\bibitem[\protect\citeauthoryear{{Popesso} et~al.,}{{Popesso} et~al.}{2023}]{Popesso23}
{Popesso} P.,  et~al., 2023, \mn@doi [\mnras] {10.1093/mnras/stac3214}, \href {https://ui.adsabs.harvard.edu/abs/2023MNRAS.519.1526P} {519, 1526}

\bibitem[\protect\citeauthoryear{{Remus} \& {Kimmig}}{{Remus} \& {Kimmig}}{2023}]{Remus23ax}
{Remus} R.-S.,  {Kimmig} L.~C.,  2023, \mn@doi [arXiv e-prints] {10.48550/arXiv.2310.16089}, \href {https://ui.adsabs.harvard.edu/abs/2023arXiv231016089R} {p. arXiv:2310.16089}

\bibitem[\protect\citeauthoryear{{Roberts}, {Parker}, {Brown}, {Joshi}, {Hlavacek-Larrondo}  \& {Wadsley}}{{Roberts} et~al.}{2019}]{Roberts19}
{Roberts} I.~D.,  {Parker} L.~C.,  {Brown} T.,  {Joshi} G.~D.,  {Hlavacek-Larrondo} J.,   {Wadsley} J.,  2019, \mn@doi [\apj] {10.3847/1538-4357/ab04f7}, \href {https://ui.adsabs.harvard.edu/abs/2019ApJ...873...42R} {873, 42}

\bibitem[\protect\citeauthoryear{{Sandles} et~al.,}{{Sandles} et~al.}{2023}]{Sandles23}
{Sandles} L.,  et~al., 2023, \mn@doi [arXiv e-prints] {10.48550/arXiv.2307.08633}, \href {https://ui.adsabs.harvard.edu/abs/2023arXiv230708633S} {p. arXiv:2307.08633}

\bibitem[\protect\citeauthoryear{{Santini} et~al.,}{{Santini} et~al.}{2017}]{Santini17}
{Santini} P.,  et~al., 2017, \mn@doi [\apj] {10.3847/1538-4357/aa8874}, \href {https://ui.adsabs.harvard.edu/abs/2017ApJ...847...76S} {847, 76}

\bibitem[\protect\citeauthoryear{{Santini} et~al.,}{{Santini} et~al.}{2019}]{Santini19}
{Santini} P.,  et~al., 2019, \mn@doi [\mnras] {10.1093/mnras/stz801}, \href {https://ui.adsabs.harvard.edu/abs/2019MNRAS.486..560S} {486, 560}

\bibitem[\protect\citeauthoryear{{Schreiber} et~al.,}{{Schreiber} et~al.}{2018a}]{Schreiber18a}
{Schreiber} C.,  et~al., 2018a, \mn@doi [\aap] {10.1051/0004-6361/201731917}, \href {https://ui.adsabs.harvard.edu/abs/2018A&A...611A..22S} {611, A22}

\bibitem[\protect\citeauthoryear{{Schreiber} et~al.,}{{Schreiber} et~al.}{2018b}]{Schreiber18b}
{Schreiber} C.,  et~al., 2018b, \mn@doi [\aap] {10.1051/0004-6361/201833070}, \href {https://ui.adsabs.harvard.edu/abs/2018A&A...618A..85S} {618, A85}

\bibitem[\protect\citeauthoryear{{Scoville} et~al.,}{{Scoville} et~al.}{2007}]{Scoville07}
{Scoville} N.,  et~al., 2007, \mn@doi [\apjs] {10.1086/516585}, \href {https://ui.adsabs.harvard.edu/abs/2007ApJS..172....1S} {172, 1}

\bibitem[\protect\citeauthoryear{{Somerville} \& {Dav{\'e}}}{{Somerville} \& {Dav{\'e}}}{2015}]{Somerville15}
{Somerville} R.~S.,  {Dav{\'e}} R.,  2015, \mn@doi [\araa] {10.1146/annurev-astro-082812-140951}, \href {https://ui.adsabs.harvard.edu/abs/2015ARA&A..53...51S} {53, 51}

\bibitem[\protect\citeauthoryear{{Speagle}, {Steinhardt}, {Capak}  \& {Silverman}}{{Speagle} et~al.}{2014}]{Speagle14}
{Speagle} J.~S.,  {Steinhardt} C.~L.,  {Capak} P.~L.,   {Silverman} J.~D.,  2014, \mn@doi [\apjs] {10.1088/0067-0049/214/2/15}, \href {https://ui.adsabs.harvard.edu/abs/2014ApJS..214...15S} {214, 15}

\bibitem[\protect\citeauthoryear{{Strait} et~al.,}{{Strait} et~al.}{2023}]{Strait23}
{Strait} V.,  et~al., 2023, \mn@doi [\apjl] {10.3847/2041-8213/acd457}, \href {https://ui.adsabs.harvard.edu/abs/2023ApJ...949L..23S} {949, L23}

\bibitem[\protect\citeauthoryear{{Suh} et~al.,}{{Suh} et~al.}{2024}]{Suh24}
{Suh} H.,  et~al., 2024, \mn@doi [arXiv e-prints] {10.48550/arXiv.2405.05333}, \href {https://ui.adsabs.harvard.edu/abs/2024arXiv240505333S} {p. arXiv:2405.05333}

\bibitem[\protect\citeauthoryear{{Tanaka} et~al.,}{{Tanaka} et~al.}{2019}]{Tanaka19}
{Tanaka} M.,  et~al., 2019, \mn@doi [\apjl] {10.3847/2041-8213/ab4ff3}, \href {https://ui.adsabs.harvard.edu/abs/2019ApJ...885L..34T} {885, L34}

\bibitem[\protect\citeauthoryear{{Thomas}, {Maraston}, {Schawinski}, {Sarzi}  \& {Silk}}{{Thomas} et~al.}{2010}]{Thomas10}
{Thomas} D.,  {Maraston} C.,  {Schawinski} K.,  {Sarzi} M.,   {Silk} J.,  2010, \mn@doi [\mnras] {10.1111/j.1365-2966.2010.16427.x}, \href {https://ui.adsabs.harvard.edu/abs/2010MNRAS.404.1775T} {404, 1775}

\bibitem[\protect\citeauthoryear{{Valentino} et~al.,}{{Valentino} et~al.}{2020}]{Valentino20}
{Valentino} F.,  et~al., 2020, \mn@doi [\apj] {10.3847/1538-4357/ab64dc}, \href {https://ui.adsabs.harvard.edu/abs/2020ApJ...889...93V} {889, 93}

\bibitem[\protect\citeauthoryear{{Valentino} et~al.,}{{Valentino} et~al.}{2023}]{Valentino23}
{Valentino} F.,  et~al., 2023, \mn@doi [\apj] {10.3847/1538-4357/acbefa}, \href {https://ui.adsabs.harvard.edu/abs/2023ApJ...947...20V} {947, 20}

\bibitem[\protect\citeauthoryear{{Weibel} et~al.,}{{Weibel} et~al.}{2024}]{Weibel24ax}
{Weibel} A.,  et~al., 2024, \mn@doi [arXiv e-prints] {10.48550/arXiv.2409.03829}, \href {https://ui.adsabs.harvard.edu/abs/2024arXiv240903829W} {p. arXiv:2409.03829}

\bibitem[\protect\citeauthoryear{{Weller}, {Pacucci}, {Ni}, {Hernquist}  \& {Park}}{{Weller} et~al.}{2024}]{Weller24ax}
{Weller} E.~J.,  {Pacucci} F.,  {Ni} Y.,  {Hernquist} L.,   {Park} M.,  2024, \mn@doi [arXiv e-prints] {10.48550/arXiv.2406.02664}, \href {https://ui.adsabs.harvard.edu/abs/2024arXiv240602664W} {p. arXiv:2406.02664}

\bibitem[\protect\citeauthoryear{{Whitaker}, {van Dokkum}, {Brammer}  \& {Franx}}{{Whitaker} et~al.}{2012}]{Whitaker12}
{Whitaker} K.~E.,  {van Dokkum} P.~G.,  {Brammer} G.,   {Franx} M.,  2012, \mn@doi [\apjl] {10.1088/2041-8205/754/2/L29}, \href {https://ui.adsabs.harvard.edu/abs/2012ApJ...754L..29W} {754, L29}

\bibitem[\protect\citeauthoryear{{Whitaker} et~al.,}{{Whitaker} et~al.}{2014}]{Whitaker14}
{Whitaker} K.~E.,  et~al., 2014, \mn@doi [\apj] {10.1088/0004-637X/795/2/104}, \href {https://ui.adsabs.harvard.edu/abs/2014ApJ...795..104W} {795, 104}

\bibitem[\protect\citeauthoryear{{Zolotov} et~al.,}{{Zolotov} et~al.}{2015}]{Zolotov15}
{Zolotov} A.,  et~al., 2015, \mn@doi [\mnras] {10.1093/mnras/stv740}, \href {https://ui.adsabs.harvard.edu/abs/2015MNRAS.450.2327Z} {450, 2327}

\bibitem[\protect\citeauthoryear{{de Graaff} et~al.,}{{de Graaff} et~al.}{2024}]{deGraaff24}
{de Graaff} A.,  et~al., 2024, \mn@doi [arXiv e-prints] {10.48550/arXiv.2404.05683}, \href {https://ui.adsabs.harvard.edu/abs/2024arXiv240405683D} {p. arXiv:2404.05683}

\bibitem[\protect\citeauthoryear{{van Dokkum} et~al.,}{{van Dokkum} et~al.}{2010}]{vanDokkum10}
{van Dokkum} P.~G.,  et~al., 2010, \mn@doi [\apj] {10.1088/0004-637X/709/2/1018}, \href {https://ui.adsabs.harvard.edu/abs/2010ApJ...709.1018V} {709, 1018}

\bibitem[\protect\citeauthoryear{{van Dokkum} et~al.,}{{van Dokkum} et~al.}{2015}]{vanDokkum15}
{van Dokkum} P.~G.,  et~al., 2015, \mn@doi [\apj] {10.1088/0004-637X/813/1/23}, \href {https://ui.adsabs.harvard.edu/abs/2015ApJ...813...23V} {813, 23}

\bibitem[\protect\citeauthoryear{{van der Wel} et~al.,}{{van der Wel} et~al.}{2014}]{vanderWel14}
{van der Wel} A.,  et~al., 2014, \mn@doi [\apj] {10.1088/0004-637X/788/1/28}, \href {https://ui.adsabs.harvard.edu/abs/2014ApJ...788...28V} {788, 28}

\makeatother
\end{thebibliography}






\bsp	
\label{lastpage}
\end{document}